\documentclass[12pt,preprint2]{aastex}
\usepackage{emulateapj5,apjfonts}
\usepackage{graphicx}% Include figure files
\usepackage{amssymb} % Define \lesssim and \gtrsim
\usepackage{bm}      % Bold math
\usepackage{onecolfloat}

\newcommand{\be}{\begin{eqnarray}}      \newcommand{\ee}{\end{eqnarray}}   
\newcommand{\ba}{\begin{array}}         \newcommand{\ea}{\end{array}}   
\newcommand{\lf}{\left}     \newcommand{\rt}{\right}
\newcommand{\fr}{\frac}   
\newcommand{\pd}{\partial}

\newcommand{\g}{\gamma} 
\newcommand{\di}{\iota}

\newcommand{\sg}{_{\gamma}}
\newcommand{\de}{_{\rm de}}
\newcommand{\snu}{_{\nu}}           
\newcommand{\eff}{_{\rm eff}}
\newcommand{\dec}{_{\rm dec}}
\newcommand{\rec}{_{\rm rec}}
\newcommand{\eq}{_{\rm eq}}

\newcommand{\Ha}{\mathcal H}
\newcommand{\lagr}{\mathcal L}
\newcommand{\Nb}{\nabla} 
\newcommand{\Nbi}{\nabla_{\!i}}    
\newcommand{\bk}{{\bf k}}
\newcommand{\bx}{{\bf x}}

\newcommand{\bn}{{\bf n}}
\newcommand{\const}{{\rm const}}

\newcommand{\lb}[1]{\label{#1}}  
\newcommand{\rf}[1]{~(\ref{#1})}
\newcommand{\ct}[1]{\ \citep{#1}}
\newcommand{\ctt}[1]{\ \citet{#1}}
\newcommand{\nbrk}[1]{\mbox{$#1$}}

\lefthead{Sergei Bashinsky}
\righthead{Mapping the Dark Kinetics}

\begin{document}


\def\head{
%  \vbox to 0pt{\vss
%                    \hbox to 0pt{\hskip 440pt\rm LA-UR-07-2136\hss}
%                   \vskip 25pt}

\title{Mapping Cosmological Observables \\
       to the Dark Kinetics}


\author{Sergei Bashinsky}
\affil{Theoretical Division, T-8, 
       Los Alamos National Laboratory, Los Alamos, NM 87545}
\email{sergeib@lanl.gov}
\date{July 5, 2007}

%% affiliation information with \ataffiltext, with one command per each
%% \altaffiltext{1}{Visiting Astronomer, Cerro Tololo Inter-American Observatory.
%% CTIO is operated by AURA, Inc.\ under contract to the National Science
%% Foundation.}

\begin{abstract}
We study systematically which features in 
the cosmic microwave background (CMB) and large-scale structure (LSS)
probe various inhomogeneous properties of the dark sectors
(including  neutrinos, dark matter, and dark energy).
We stress, and quantify by simple formulas, 
that the primary CMB anisotropies are very susceptible 
to the gravitational potentials during horizon entry, 
less at recombination.
The CMB thus allows us to scan $\Phi+\Psi$ and the underlying dark kinetics
for all redshifts $z\sim 1 - 10^5$.
LSS, on the other hand, responds strongest to $\Phi$ at low redshifts.
~Dark perturbations are often parameterized by
the anisotropic stress and effective sound speed (stiffness).
We find that the dark anisotropic stress and stiffness
influence the visible species at the correspondingly early 
and late stages of horizon entry,
and affect stronger respectively the CMB and LSS.
The CMB yet remains essential to probing the 
stiff perturbations of light neutrinos and dark energy,
detectable only during horizon entry.  
The clustering of dark species and large propagation speed of their
inhomogeneities also map to distinctive features in the CMB and LSS.
~~Any parameterization of the signatures of dark kinetics
that assumes general relativity can effectively 
accommodate any modified gravity~(MG)
that retains the equivalence principle for the visible sectors.
This implies that formally the nonstandard structure growth or
$\Phi/\Psi$ ratio, while indicative, are not definitive MG signatures.
The definitive signatures of MG may include  
the strong dependence of the apparent dark dynamics on visible species,
its superluminality, and the nonstandard phenomenology of gravitational waves.
\end{abstract}
\keywords{cosmology: theory --- cosmic microwave background 
           --- large-scale structure of universe  --- dark energy --- modified gravity}
}



\twocolumn[\head]

\section{Introduction}
\lb{sec_intro}

Galactic and cluster dynamics,
cosmic structure, type Ia supernovae,
the cosmic microwave background (CMB),
and the primordial abundances of light elements
provide solid evidence that dark sectors
constitute a significant energy fraction of the universe 
at any accessible redshift $z \lesssim 10^{10}$.
At all corresponding cosmological epochs 
the nature of abundant dark species, 
coupled to photons and baryons only by gravitation,  
is partly or entirely uncertain.

The mainstream analyses of cosmological data usually assume 
the minimal neutrino sector,
non-interacting cold dark matter (CDM), 
and dark energy represented by a canonical scalar field (quintessence). 
These assumptions are reasonable for interpreting the available data,
yet none of them can be taken for granted.
~For example, new light weakly interacting particles commonly appear
in high-energy models.
In some models, even the standard neutrinos 
recouple to each other or to additional light fields 
\ct{Chacko:2003dt,Chacko:2004cz,Beacom:2004yd,Okui:2004xn,Grossman:2005ej}
at redshifts at which the decoupled component of radiation gravitationally affects 
the CMB and cosmic structure
\citep{HuSugSmall96,BS04,Hannestad:2004qu,Bell:2005dr}.
~Various alternatives to cold dark matter
have been suggested as well.
These include warm dark matter\ct{Blumenthal:1982mv,Olive:1981ak}, 
self-interacting dark matter\ct{Carlson_et_al_92,deLaix:1995vi,Spergel:1999mh}, 
or modified gravity\ct{Milgrom83,Bekenstein04,Skordis:2005xk,Dodelson:2006zt}.
The viability of such scenarios remains an intriguing question.
~Quintessence models are convenient for 
quantitatively constraining dark energy parameters by data. 
Yet quintessence is not readily motivated by particle physics, 
where it is difficult to naturally achieve
the required shallowness of the field potential.
On the other hand, 
many alternatives have been proposed whose inhomogeneous kinetics,
and hence cosmological signatures,
{\it cannot be mimicked by quintessence with any background\/} equation of state~$w(z)$.
\citep[For a comprehensive review of dark energy models 
see, e.g.,][]{Copeland:2006wr}.

Fortunately, cosmological observations themselves
can test these assumptions by revealing not only the dark species'
mean density and pressure but also the kinetics of their inhomogeneities.
The goal of this paper is to map
various inhomogeneous kinetic properties of the dark sectors
(or deviations from Einstein gravity)
to the observable characteristics of CMB and cosmic structure.
Dark species influence the visible matter
by affecting both the background expansion and metric perturbations.
Of the two mechanisms, the perturbations, 
albeit demanding better statistics for useful constraints,
encode many more independent clues about the dark universe 
by offering {\it new information at every spatial scale\/}~$k$.
The following three examples show the importance of this information, 
absent in the background equation of state~$w(z)$.

\subsection{Examples of the value of dark perturbations}

\subsubsubsection{Nature of dark energy}

The first example is the most challenging problem in today's cosmology---the nature of dark energy.
The constraints on the dark energy background equation of state $w\equiv p\de/\rho\de$ 
are tightening around the value $-1$, consistent with a cosmological constant.
Analyses that combine the current data from the CMB, large scale structure (LSS), 
Lyman-$\alpha$ forest, and supernovae, 
already constrain the deviation of $w$ from $-1$ 
for flat models better than to $10\%$
\citep[][and others.]{WMAP3Spergel,SelSlosMcD06,TegmLRG06} 
Whether or not future observations
continue to converge on $w=-1$,
the dynamics of perturbations will be crucial in elucidating 
the nature of cosmic acceleration.

Even if $w(z)\equiv -1$ at low redshifts,
this does not necessarily imply a cosmological constant.
Certain models \cite[e.g.][]{MaVaN2Fardon05}
predict that $w\equiv -1$ at the present epoch, 
yet the dynamics of perturbations differs at high redshifts.
Contrary to common belief, it is even conceivable that $w\approx -1$
yet low-redshift perturbations deviate from those in the $\Lambda$CDM model.
On the other hand, if $w\not=-1$ then
perturbations of dark energy in the inhomogeneous metric are unavoidable.
(Even if dark energy appears unperturbed in one spacetime slicing,
 its perturbations for $w\not=-1$ are necessarily nonzero in any different slicing.)
Exploring the perturbations' properties, 
specifically the properties considered in this paper,
will then become pivotal to establishing the nature of dark energy.

\subsubsubsection{Density of dark radiation}

The current observations of CMB temperature and polarization,
including the WMAP 3-year results\ct{WMAP3Hinshaw,WMAP3Page}, 
when combined with the SDSS galaxy power spectrum\ct{Tegmark:2003uf,Eisenstein:2005su},  
with or without Ly-$\alpha$ forest data, prefer enhanced  neutrino density.
\ctt{SelSlosMcD06} report $N_\nu=5.3^{+2.1}_{-1.7}$ at $2\,\sigma$,
disfavoring the standard value $N_\nu=3.04$ at $2.4\,\sigma$.
The WMAP team in its latest analysis\ct{WMAP3Spergel} likewise concludes that 
from the WMAP3 and SDSS data $N_\nu=7.1^{+4.1}_{-3.5}$.
A similar preference for high $N_\nu$ by the WMAP3 and SDSS data
is seen by\ctt{Cirelli:2006kt},  
%and\ctt{Friedland:2007vv}
although not by\ctt{Hannestad:2006mi}.

While neutrinos noticeably speed-up the background expansion in the radiation era,
by itself this leads to 
almost no observable cosmological signatures\ct{HuEisenTegmWhite98,BS04,B05Trieste},
given the freedom of a compensating adjustment of
matter density $\Omega_m h^2$\ct{Bowen02} 
and primordial helium fraction~$Y_p$\ct{BS04}.
 
This degeneracy is broken by 
the differences in the evolution of 
streaming neutrino perturbations 
and of the acoustic waves in the photon-baryon fluid.
It is also broken by independent constraints on the helium fraction.
Various signatures of neutrinos that then appear
remain partly degenerate with nuisance parameters,
such as the dark energy equation of state
or the normalization and tilt of the primordial power.
Yet these signatures and the constraints on $N_\nu$
with the correspondingly different experiments 
have rather incomparable degree of robustness\ct{BS04}.

The preference of WMAP3+SDSS data for an enhanced neutrino density
could be due to the physical excess of matter power over CMB power, 
expected for higher density of freely streaming particles\ct{HuSugSmall96,BS04}.
Yet other explanations, e.g., incomplete treatment of recombination 
or insufficient accuracy of bias modeling, also cannot be definitely excluded.

In addition to increasing the ratio of matter to CMB power,
freely streaming neutrinos shift the phase of CMB acoustic oscillations\ct{BS04}.
The phase-shift signature is very robust to systematic uncertainties
and will become the primary discriminative mechanism 
for the near-future tight CMB constraints on neutrino 
density\ct{Lopez:1998aq,Bowen02,BS04,Perotto:2006rj},
e.g. $\sigma(N_\nu)\sim 0.2-0.3$ expected from
the Planck mission\footnote{http://www.rssd.esa.int/Planck}.
It can provide decisive evidence for the excess
of streaming relativistic particles or rule this possibility out.

~

\subsubsubsection{Nature of dark radiation}

If there is a true excess of energy density in the radiation era,
at least three alternatives for non-standard dark radiation are possible:
relic decoupled particles, 
self-interacting particles\ct{Chacko:2003dt,Chacko:2004cz,Beacom:2004yd,Hannestad:2004qu},
or a tracking classical field\ct{RatraPeebles88,FerreiraJoyce97,ZlatevStein_tracking98}. 
The perturbations of dark radiation propagate differently in all of these scenarios. 
As follows from this work, this in principle allows their experimental discrimination.

\subsection{Questions not answered by black-box computations}

For a particular cosmological model it is generally straightforward to calculate 
linear power spectra and transfer functions with standard codes,
if necessary, modified to include new dynamics.
Despite this, numerical calculations have limited usefulness
for exploring the signatures of new physics:   First, for typical models, 
with close to ten unknown nuisance parameters,
it is often intricate to establish numerically 
which of the observable signatures of the new physics cannot be
compensated by parameter adjustments.
While only such signatures are the true discriminators of the physics in question,
they may be tiny and easily overlooked among large yet degenerate effects.
Moreover, for every extended parameterization of the nuisance effects
as well as for any additional constraining experiment,
a new numerical analysis is required.

Equally importantly, a numerical black-box computation 
does not answer which aspects of a model
are responsible for its observable signatures.
This obscures the separation of the physical facts  
that are backed by observations from the models' features that are
believed true yet are untested.

\subsection{Approach, outline, and conventions}

The approach of this paper is to explicitly track
the evolution of gravitationally coupled inhomogeneities of 
the visible and dark species.
This allows us to identify which observables
are affected by which of the various properties of the dark kinetics.
Importantly, this approach also reveals the mechanisms behind 
the sensitivity of cosmological observables to various dark properties.
Establishing the mechanism allows us to judge more intelligently 
the robustness of cosmological constraints,
particularly, to know when this mechanism may fail or produce a different outcome. 

As we will see soon, there is an important subtlety in performing such identifications.
To decipher the signatures of the dark kinetics, it is essential to address   
the gravitational interaction of perturbations during {\it horizon entry\/}. 
Then and only then perturbations of all abundant dark species 
are gravitationally imprinted on the visible species without suppression. 
The suppression of  the impact of the species' energy overdensity 
$\delta\rho/\rho$ on the metric on subhorizon scales 
is evident from the Poisson equation
\be
\Phi\,=\,-k^{-2} 4\pi G a^2\,\delta\rho \sim (\Ha/k)^2\,\delta\rho/\rho,
\lb{Pois_subhor}
\ee
where $\bm{k}$ is the (comoving) wavevector of a perturbation mode
and $\Ha$ is the expansion rate (in conformal time).
The impact of the species' velocities 
is suppressed even more [by $(\Ha/k)^3$, c.f.\ eq.\rf{Poiss_my}.]
At horizon entry, on the other hand,
the factor $\Ha/k$ approaches unity and does not cause suppression.
In particular, only during the horizon entry are the visible species influenced by
the perturbations of dark radiation and dark energy, 
for which the Jeans length is expected to be close to the horizon size.

Numerous authors \citep[e.g.,][]{PressVishniac80,Bard80,MaBert95} have pointed out
considerable freedom in representing the inhomogeneous evolution 
on the scales of the order of and exceeding the Hubble scale.
This freedom is due to, literally, an infinite number of possibilities for coordinate gauges,
as well as for the variables (even the gauge-invariant ones)
that can parameterize large-scale perturbations.   
In principle, this descriptional freedom could introduce ambiguity
in relating observable features with specific physical mechanisms
that operate on large scales.
Moreover, such relations indeed differ
among the authors who use apparently dissimilar while formally equivalent descriptions---
differ substantially for some pronounced and important for constraints features.

We argue in Sec.~\ref{sec_uniq} that, 
as far as the observable impact of dark dynamics is concerned,
there is little room for ambiguities.
A well defined distinction can be drawn between an apparent
connection ``dynamical cause $\to$ observable effect''  
that is a descriptional artifact or
that is an objective causal relation.
We will also see that in certain formalisms,
which exist and can be distinguished by simple criteria,
the physical microscopic properties that characterize species 
at a particular time influence 
the apparent large-scale perturbations {\it at the same time\/}.
Such formalisms markedly simplify mapping of the characteristics of 
cosmological observables 
to the responsible microscopic dark kinetics.

In the subsequent sections
we do such a mapping for the CMB temperature power spectrum
and LSS transfer functions.
In Sec.~\ref{sec_probes} we discuss the general features of 
the primary probes of inhomogeneous cosmological dynamics. 
We also give dynamical equations to be used further to quantify 
the CMB and matter response to dark parameters.
In Sec.~\ref{sec_dynamics} we review various general parameterizations 
of potentially accessible information about the properties of dark sectors.
We consider the parameterizations of the metric, 
of dark densities and stresses,
and of internal dark dynamics.
Sec.~\ref{sec_impacts} is central to our study. 
In this section we identify the characteristics of the CMB and cosmic structure 
that reveal such general properties of dark species as
anisotropic stress, stiffness, clustering, and propagation of inhomogeneities.
In Sec.~\ref{sec_ModGrav} we study the specifics of cosmologies 
with modified gravity and discuss how modified gravity
can be distinguished observationally.
Our main results are summarized in Sec.~\ref{concl} and its Table~\ref{tab_sum}.

Throughout the paper, distances will be measured in comoving units.
Evolution will usually be described in conformal time $d\tau=dt/a$, 
where $a$~is the cosmological scale factor
and $dt$ is proper background time.
Overdots denote derivatives with respect to 
the conformal time, and $\Ha\equiv \dot a/a = Ha$
gives the Hubble expansion rate in conformal time.

\section{Is mapping of large-scale dynamics to observables unambiguous?}
\lb{sec_uniq}

Except for Sec.~\ref{sec_ModGrav},
we will consider models in which dark and visible sectors
couple by standard Einstein gravity.
These models correspond to a local action $S=\int d^4x\,\sqrt{-g}\,\lagr$
with $\lagr=\lagr_{\rm dark}+\lagr_{\rm vis}+\lagr_{\rm grav}$,
where the dark degrees of freedom are represented by 
the Lagrangian density $\lagr_{\rm dark}$,
the visible ones by $\lagr_{\rm vis}$, and 
the gravitational ones by $\lagr_{\rm grav}=(16\pi G)^{-1} R$.
In agreement with observations, we assume
that the large-scale evolution can be presented as
a mildly perturbed Friedmann-Robertson-Walker (FRW) expansion.
For this section only, we clock the evolution by the
number of $e$-foldings $N\equiv\ln a$, tractable through 
local matter density or CMB temperature.

Let $[N_1, N_2]$ be an evolution interval in the past.
Consider a class,~$M_{12}$, of all conceivable models 
with same $\lagr_{\rm vis}$ that: 
\begin{itemize}
\item[(i)] evolve identically for $N<N_1$;
\item[(ii)] during the interval $[N_1, N_2]$
possibly differ in the microscopic laws of their dark, but not visible, dynamics; 
and
\item[(iii)] by $N=N_2$ have identical distribution of the dark species
(in some non-degenerate measure) among all~$M_{12}$ models.
\end{itemize}
By $N=N_2$, the distributions of the visible species among these models will generally differ: 
The visible species would generally be gravitationally affected earlier by 
the model-specific evolution in the dark sectors.\footnote{
  Even the dark species in the $M_{12}$ models may not be perturbed identically
  at $N_2$ in terms of every measure.
  On large scales the quantification of dark perturbations by a different measure
  can depend on the perturbations of visible species and the metric.
  We require only that the dark perturbations are the same in at least one 
  non-degenerate measure.
}

Given a quantity $P$ that characterizes internal dark dynamics and an observable~$O$,
the following may be true: 
Regardless of the value of $P$ during the interval $\Delta N_{12}\equiv [N_1, N_2]$, all the models 
in the above class~$M_{12}$ have the same observed value of~$O$.
Then it is natural to say that the observable~$O$ is {\it not sensitive\/} to  
the considered dynamical property~$P$ in the interval $\Delta N_{12}$.

In linear order, this applies to the effects of 
any dark property~$P$ on any observable~$O$ that depends only on the perturbations 
that were superhorizon during $\Delta N_{12}$.
\ctt{zeta_a} showed that perturbations in
several gravitationally coupled fluids can be described by
variables which during superhorizon evolution (for non-decaying modes) 
are time-independent in each of the individual fluids.\footnote{
  \ctt{LindeMukh_curvaton05} noted that 
  under certain conditions, which may exist, e.g., during reheating,
  {\it gravitational\/} decays
  of species into another type of species may mix 
  the superhorizon perturbations in different sectors.
  Such decays are, nevertheless, negligible
  for the post-BBN evolution studied in this paper.
}
This result can be extended beyond multifluid models
to any sector that is perturbed (internally) adiabatically,
i.e., is  homogeneous in at least some coordinates\ct{BS04}.
In other words, linear perturbations
encode no information about the microscopic properties 
that characterized the dark universe
when these perturbation were superhorizon.

The abundances and kinetic properties 
that dark species have {\it since horizon entry\/}
do influence the observed CMB or matter perturbations.
However, because of the aforementioned freedom in representing 
the evolution of large-scale inhomogeneities,
in many formalisms
(including the most popular ones),
the properties characterizing the dark species at horizon entry 
affect the apparent perturbations of the CMB or matter long before or after the entry.
This can mislead one into viewing an observable feature 
as a probe of an entirely unrelated epoch and/or physical process.
This source of existing and potential new misassignments
can be eliminated systematically and naturally as follows.

The impact of local dark dynamics on perturbations 
of visible species will appear concurrent 
with the underlying microscopic dark physics in any formalism
which has the following two properties\ct{B06}:

~

\begin{itemize}
\item[I.] Perturbations are frozen on superhorizon scales.
\item[II.] Perturbations evolve by the equations of the FRW metric whenever the geometry
is unperturbed.
\end{itemize}
We imply that the description of gravity in these formalisms reduces to the Newtonian one
in the weakly perturbed metric on subhorizon scales.
This is easily achieved, e.g., by parameterizing metric inhomogeneities
by the gravitational potentials of the Newtonian gauge \citep{Mukh_Rept92,MaBert95}.

Then the apparent linear impact of dark species on observables will be found to be identical
in any description with properties I~and~II:
This impact is practically unambiguous for the evolution 
of perturbation modes after horizon entry;
the apparent perturbations do not evolve at all prior to the entry
by condition~I.
Therefore, these descriptions could differ only by 
the changes of perturbations during horizon entry.
It is straightforward to argue\ct{B06} that  
there are no such differences in linear theory among the formalisms that satisfy I~and~II.

One such particular, natural and simple, formulation 
of the full linearly perturbed Einstein-Boltzmann cosmological dynamics
was described in\ctt{B06}.
This formulation is based on {\it canonical\/} coordinate variables.
The canonical variables for perturbations
of phase-space distributions or radiation intensity were considered
in the past \citep[e.g.,][]{Chandrasekhar_book60,Durrer_CMB01}
but have not become mainstream in the present cosmology.
In addition to the demonstrated more direct connection of large-scale evolution
to microscopic kinetics of cosmological species,
this formulation has several technical advantages 
over popular alternatives, 
most of which consider perturbations of proper quantities.
It is used in the analysis that follows.

\section{Probes}
\label{sec_probes}

Following the standard route, 
we separate the effects of dark sectors on the metric
into the contributions to the background geometry and 
to metric inhomogeneities.
The background geometry is fully described 
by the Hubble expansion rate as a function of redshift
and by possible global curvature.
It is constrained by a variety of ``geometrical'' probes,
e.g., the luminosity-redshift relation for type Ia supernovae
or the angular size of the acoustic features 
in the power spectra of the CMB and matter. 
In this paper we focus on extracting 
the potentially much richer information
about the dark dynamics that is contained in the metric inhomogeneities.
This information is to be inferred from
the imprints of the metric inhomogeneities on such various observables
as the CMB, galaxy distributions, lensing shear,
quasar absorption spectra, etc.

The primary probes of dark perturbations can be 
broadly classified as either ``light'' or ``matter'':
The ``light'' probing the metric along trajectories which are close to 
null geodesics ($ds^2\simeq 0$, e.g., CMB photons);
the ``matter'' moving almost with the Hubble flow 
with non-relativistic peculiar velocities 
($|d\bm{x}/d\tau|\ll 1$, e.g., 
galaxies or Ly-$\alpha$ clouds).
These two classes of probes,
considered in Secs.~\ref{sec_CMB} and~\ref{sec_matter} respectively,
provide information that is complimentary in several respects,
discussed in details in subsequent sections.

Linear theory adequately describes
the inhomogeneous dynamics during horizon entry,
when physical quantities are perturbed by about $0.001\%$,
and long after the entry.
Then the signatures of the dark kinetics are
encoded in the linear transfer functions [alternatively, 
in Green's functions\ct{Magueijo92,Baccigalupi98,Baccigal2,BB_PRL,BB02}],
which are the essential constituents 
of ``dynamical'' cosmological observables, such as
power spectra, luminosity functions, etc.
To be specific (though general within the linear regime), 
in this section we consider an individual perturbation mode with a comoving wavevector~$\bk$.

\subsection{CMB}
\label{sec_CMB}

After horizon entry until photon decoupling around hydrogen recombination,
a perturbation mode in the photon-baryon plasma 
undergoes acoustic oscillations.
All the memory of the gravitational impact at the mode's entry 
is then retained only in its oscillation amplitude~$A(\bk)$
and its temporal phase shift~$\varphi(k)$.
These quantities map into
the heights and phase of the acoustic peaks
in the observable CMB angular power spectra~$C_l$,
with a rough $\ell\leftrightarrow k$ correspondence
$\ell\sim k\,r\dec$, where $r\dec$ is the angular diameter distance
to the CMB decoupling surface.
[In the standard $\Lambda$CDM model $r\dec=\tau(z\dec{\sim}1100)\simeq 14\,$Gpc.]

Our goal is to establish how $A(\bk)$ and $\varphi(k)$ are affected by 
the gravitational impact of perturbations in various species.
As noted in the introduction, this task involves a subtlety that without
proper care can lead to erroneous conclusions.
Although the notion of density or temperature perturbation is  
uniquely defined in the FRW geometry,
ambiguities arise on large scales in the perturbed metric.
We argued in Sec.~\ref{sec_uniq}
that this complication is uniquely resolved in linear theory,
where switching off a microscopic effect of interest 
results in the same change of observables
regardless of other local properties of the various species.
We can then calculate the observables
in the models with the effect ``on'' and ``off'' 
and identify the {\it signature of the effect\/} with the difference.
Being interested in the kinetics of perturbations,
we compare the models with identical background expansion $H(z)$.
For the compared models,
we assume identical initial conditions,
i.e., identical dynamics during the inflationary epoch, 
when the superhorizon perturbations likely were generated.

We describe the perturbative evolution by
a formalism detailed in\ctt{B06}.
Then the gravitational forcing of perturbations
appears concurrent with the responsible local interactions and
the above unambiguous causal relations are manifest.
In this formalism 
the general-relativistic generalization of particle overdensity
is $d\equiv \delta n_{\rm coo}/n_{\rm coo}$,  
a perturbation of particle number density per {\it coordinate\/} volume.

When the CMB is tightly coupled to electrons by Thompson scattering,
its (coordinate) overdensity evolves as
\be
\textstyle
\ddot d\sg + \left(\fr{R_b}{1+R_b}\,\Ha + 2\tau_d k^2\right)\dot d\sg
                                                    + c_s^2k^2(d\sg-D) = 0
\lb{dot_gk}
\ee
\citep{BS04}. Here and below $R_b\equiv 3\rho_b/(4\rho_{\g})$,
the diffusion time $\tau_d=[1-\fr{14}{15}(1+R_b)^{-1}
 +(1+R_b)^{-2}]/(6an_e\sigma_{\!\rm Thomp})$\ct{Kaiser83}
determines the Silk damping\ct{Silk:1967kq}, 
and $c_s^2=[3(1+R_b)]^{-1}$ gives the sound speed in the photon-baryon plasma.
Only scalar perturbations are considered.\footnote{
 The connection between dark dynamics and its observable signatures
 is relatively straightforward in the tensor sector,
 where gauge ambiguities are absent\ct{Bard80,KS84}.
 Vector perturbations, even if primordially generated, 
 are not expected to survive superhorizon evolution\ct{Bard80,KS84}. 
 If necessary, they can be analyzed similarly to the scalar perturbations.
}
The gravitational driving of the CMB modes in eq.\rf{dot_gk}
is mediated through an instantaneous equilibrium value~$D$
of photon overdensity.
In the Newtonian gauge \citep{Mukh_Rept92,MaBert95} 
\be
ds^2=a^2\left[-(1+2\Phi)d\tau^2+(1-2\Psi)d\bm{x}^2\right],
\lb{Newt_gauge}
\ee
the driving term equals 
\be
D(\tau,k)= -3(\Phi+\Psi+R_b\Phi).
\lb{D_def}
\ee

In the radiation era, when $R_b\ll 1$, 
the driving of the CMB by dark perturbations {\it on all scales\/}
is controlled only by the sum $\Phi+\Psi$.
This also applies to the CMB photons on all scales
after their decoupling from baryons,
c.f.\ the equation for the evolution of CMB intensity\rf{dot_dig_C}.
\ctt{Durrer_pert93} argued that this is natural for the relativistic particles,
whose dynamics is conformally invariant, and therefore should be 
sensitive only to the Weyl part of the curvature tensor.
Indeed, the scalar part the Weyl tensor is fully specified  by 
$\Phi+\Psi$\ct{Durrer_pert93}. 
In the decoupled case, the driving by $\Phi+\Psi$
is well known as the integrated Sachs-Wolfe effect (ISW)\ct{SachsWolfe67}.

The full linear dynamics of 
partially polarized CMB photons and baryons \cite[e.g.,][]{MaBert95}
is also straightforward to formulate so that 
the changes of the apparent inhomogeneities 
are concurrent with the responsible local dark properties\ct{B06}.
In particular, the perturbation 
of the polarization-summed CMB intensity
corresponds to its canonical variable $\di(x^\mu,n_i)\equiv I\sg/\bar I\sg-1$
that evolves according to a transport equation
\citep[][also presenting a fully nonlinear treatment]{Durrer:1990mk,SB_tensors05}
\be
\dot \di + n_i\Nbi \di 
  = -4 n_i\Nbi(\Phi+\Psi) + C_T,
\lb{dot_dig_C}
\ee
where $C_T$ is the Thompson collision term.

\subsection{Matter}
\label{sec_matter}

The evolution of linear scalar perturbations of 
cold dark matter (CDM) on all scales 
and, after decoupling from the CMB, of baryons on large scales 
is governed by the conservation and Euler equations
\be
\dot d_c + \pd_i v^i_c = 0, \qquad
\dot v^i_c + \Ha v^i_c = -\pd_i\Phi.  
\lb{dot_c}
\ee
Their solution for a Fourier mode with non-singular (inflationary) initial conditions is 
\be
d_c = d_{c,\,\rm in} 
     - k^2\int_0^\tau u_c\, d\tau, \qquad
u_c = \fr1a\int_0^{\tau}(a\Phi)\,d\tau.
\lb{d_c_sol}
\ee
The quantity $u_c$ that appears in these equations is physically the CDM velocity potential:
$v^i_c=-\pd_i u_c$.

Similarly to the CMB modes,
the matter perturbations carry the memory 
of the metric inhomogeneities at the horizon entry 
in two independent functions:
the {\it overdensity change\/} and {\it velocity boost\/} 
that were generated during the entry.
After horizon entry, the excess of CDM velocity,
decaying as~$1/a$, yet not vanishing instantaneously,
continues to enhance matter clumping.
Hence, the implications of the velocity boost 
for the observed $\delta\rho_m/\rho_m$
are generally more important than 
the initial overdensity enhancement\ct{HuSugSmall96}.

Eqs.\rf{d_c_sol} show that matter perturbations 
are driven by the potential~$\Phi$ alone {\it on all scales\/}\ct{BS04}.
This also has a natural explanation.
In the Newtonian gauge, the point particles of a mass $m$ are described by the Lagrangian
$L=-m\sqrt{-ds^2}/d\tau=-ma\sqrt{[1+2\Phi({\bf x})]-[1-2\Psi({\bf x})]\,\dot{\bf x}^2}$.
The corresponding contribution of $\Psi$ is negligible for nonrelativistic particles,
when $L\simeq ma\,[\dot{\bf x}^2/2-\Phi({\bf x})]\,+\,$const.
Moreover, the initial (superhorizon) perturbations of particle's coordinates ${\bf x}$
determine the initial value of the {\it coordinate\/} overdensity $d_c$ 
also independently of $\Psi$.

Since the CMB, on the other hand, is
affected by the sum~$\Phi+\Psi$,
this is one of the respects in which 
matter and the CMB provide complimentary information about dark inhomogeneities.
A combined analysis of both probes is particularly useful 
for testing a deviation
of~$\Psi/\Phi$ from unity,
produced by freely streaming species \cite[e.g.,][]{MaBert95}
or generally expected in modified gravity\ct{Bert06,Dodelson:2006zt}.

\section{Parameterizing dark dynamics}
\label{sec_dynamics}

In this section we review several general parameterizations 
of the potentially observable properties of the dark universe.
We start from the spacetime metric, 
connected most directly to observational probes.
Then, assuming validity of the Einstein equations, 
we match the metric characteristics to a general description of dark densities and stresses.
Finally, we consider a parameterization
of the potentially measurable properties of 
dark internal dynamics.
The observational probes of these parameters 
will be studied in Sec.~\ref{sec_impacts}.
The status of these parameterizations in cosmologies with modified gravity
will be addressed in Sec.~\ref{sec_ModGrav}.

\subsection{Metric}
\label{sec_dynamics_metric}

Provided the visible matter couples covariantly and minimally to a certain metric tensor $g_{\mu\nu}$,
the observable impact of dark sectors or modified gravity alike
can be parameterized by a single function of redshift~$z$,
to describe the FRW background,
and by two independent functions of $z$ and~$\bk$,
to specify scalar metric perturbations.
For example, the background can be quantified by its expansion rate~$\Ha(z)$
(with its possible spatial curvature),
and the metric perturbations by
the Newtonian-gauge potentials $\Phi(z,\bk)$ and $\Psi(z,\bk)$.

\subsection{Densities and stresses}

For a more direct link to the internal properties
of the dark sectors, it may be worthwhile to parameterize
their contribution to the energy-momentum tensor
$T^{\mu\nu}=\sum_a T^{\mu\nu}_a$.
If gravity is described by general relativity
then $T^{\mu\nu}$~determines the geometrical quantities
$\{\Ha$, $\Phi$, $\Psi\}$ through the Einstein equations.
In this view, the background may be specified
by the average energy densities 
of the species $\rho_a(z)\equiv -\bar T^0_{0\,(a)}$.
The scalar inhomogeneities of the metric
are determined from the perturbations of the energy densities,
sourcing the curvature potential~$\Psi$,
and from anisotropic stress,
generating the splitting $\Phi-\Psi$.
The corresponding formulas are given next.

After\ctt{Hu_GDM98}, we will use proper particle number overdensity
$\delta^{(c)}_a(z,\bk) \equiv \delta\rho_a^{(c)}/(\rho_a+p_a)$
in the {\it comoving\/} gauge\footnote{
  The comoving gauge conditions for scalar perturbations
  are zero total momentum density ($T^0_i \equiv 0$) 
  and vanishing shear of the spatial metric ($g_{ij} \propto \delta_{ij}$).
}.
In any other gauge
\be
\delta^{(c)}_a = \fr{\delta\rho_a}{\rho_a+p_a}+3\Ha u_a,
\lb{delta^c_def}
\ee 
where $v^i_a= -\pd_i u_a$.
Specifically, $\delta^{(c)}_a$ can be expressed 
in terms of the dynamical coordinate overdensities~$d_a$,
used in Sec.~\ref{sec_probes}, as
\be
\delta^{(c)}_a = d_a + 3\Ha u_a + 3\Psi.
\lb{delta^c_Newt}
\ee

The comoving-gauge overdensity $\delta^{(c)}_a$ is convenient for two reasons.
First, it directly sources 
the Newtonian potential~$\Psi$\ct{Bard80} 
\be
\Nb^2\Psi = 4\pi Ga^2\sum_a(\rho_a+p_a)\delta^{(c)}_a,
\lb{Poiss_tradit}
\ee
where $\Nb^2\equiv {}^{(3)}\!\bar g^{ij}\pd_i\pd_j$ is the spatial Laplacian
(from now on the background is assumed flat).
Second, the generalized pressure gradient 
in the Euler equation
[the term $- \pd_i f_p$ in eq.\rf{dot_v}]
in the comoving gauge takes its usual special-relativistic form
[$f_p = \delta p^{(c)}/(\rho+p)$].

Nevertheless, we view $\delta^{(c)}_a$
only as a useful linear combination\rf{delta^c_Newt},
rather than an independent dynamical variable: 
The evolution of~$\delta^{(c)}_a$, 
unlike that of~$d_a$, is highly nonlocal.
Beyond the horizon~$\delta^{(c)}_a$, unlike the frozen $d_a$, 
is affected by all species,
in addition to the studied dark species~$a$.
In terms of the dynamical variables $d_a$ and $u_a$
the Poisson equation\rf{Poiss_tradit} reads
\be
\left(-3+\fr{\Nb^2}{\gamma}\right)\Psi = 
             \sum_a x_a\,(d_a + 3\Ha u_a),
\lb{Poiss_my}
\ee
with
\be
\gamma \equiv 4\pi Ga^2(\rho+p) = \frac32\,(1+w)\Ha^2,
~~~{\rm and}~~~ x_a\equiv \frac{\rho_a+p_a}{\rho+p}.
\lb{Poiss_my_notat}
\ee
Unlike eq.\rf{Poiss_tradit}, the generalized Poisson equation  
in the form\rf{Poiss_my} in the superhorizon limit
is explicitly non-singular and its solution $\Psi \approx -\fr13(d + 3\Ha u)$
even becomes local.

We describe the scalar component of the 
anisotropic stress $\Sigma^i_j\equiv T^i_j-\fr13\delta^i_jT^k_k$
by a potential~$\sigma(z,\bk)$ as
\be
(\pd_i\pd_j-\fr13\delta^i_j\Nb^2)\sigma_a
  \equiv {\Sigma^i_{j\,(a)}\over(\rho_a+p_a)}.
\lb{sigma_def}
\ee
The corresponding gravitational equation is\ct{Bard80,KS84,MaBert95}:
\be
\Phi=\Psi-8\pi Ga^2\sum_a(\rho_a+p_a)\sigma_a.
\lb{Psi-Phi}
\ee

Thus the densities and stresses of the background and scalar perturbations
may be parameterized by a set
$\{ \rho_a(z)$, $\delta^{(c)}_a(z,\bk)$, $\sigma_a(z,\bk)\}$
for each sector~$a$.
Both $\delta^{(c)}_a$ and $\sigma_a$,
as defined by eqs.\rf{delta^c_def} and\rf{sigma_def},
are gauge-invariant quantities.
Neither of them, though, can be measured locally.

\begin{figure*}[t]
\centerline{\sc\footnotesize \qquad CMB modes, Radiation era}
\centerline{\includegraphics[width=12cm]{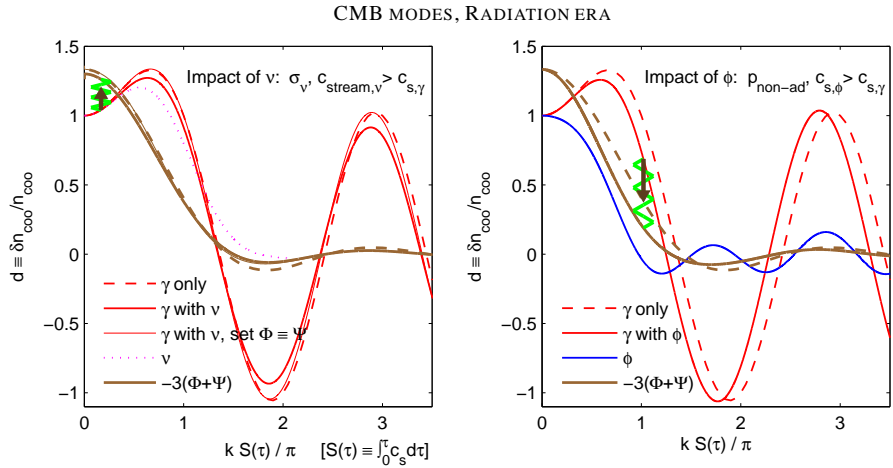}}
\caption{The evolution of CMB overdensity~$d\sg(\tau,k)$
(oscillating, red curves) and the corresponding gravitational driving term~$D=-3(\Phi+\Psi)$ 
(monotonically falling, brown curves) 
in the radiation era for $R_b\ll 1$ and negligible photon diffusion.  
The gravitational driving of the CMB [eq.\rf{dot_gk}]
is illustrated with a spring that connects the $d_\gamma$ and $D$ curves.
The panels demonstrate the impact of perturbations
in decoupled neutrinos (left) and quintessence (right).
---\,Left panel shows $d_\gamma$ and $D$ for:
tightly-coupled photons only (dashed),
the standard model with three neutrinos (wider solid),
and the same model after switching off neutrino anisotropic stress 
as a source of $\Phi-\Psi$ (thinner solid).
---\,Right panel displays: the earlier photon-only model (dashed) and
a fictitious model where, as in the standard model, $59\%$ of energy density is in photons
but the remaining $41\%$ is now in a tracking quintessence~$\phi$, with $w_\phi=1/3$ (solid).
}
\label{fig_CMB_rad}
\end{figure*}

\subsection{Dynamical properties}
\label{sec_param_dyn}

As a step further, we can try to parameterize 
the {\it dynamics\/} of the dark densities and stresses.
One such useful and popular parameterization has been suggested by \citet{Hu_GDM98}.
He described the dynamics of the background
by the conventional ``equation of state'' $w_a(z)\equiv p_a/\rho_a$,
and the dynamics of perturbations by
the effective sound speed 
\be
c^2_{{\rm eff},\,a}(z,k)\equiv \delta p_a^{(c)}(k)/\delta\rho_a^{(c)}(k).
\lb{c_eff_def}
\ee
We call this quantity ``stiffness'' throughout the paper 
in order to distinguish it from the physical velocity of perturbation propagation $c_p$,
Sec.~\ref{sec_speed}, which may differ from $c_{\rm eff}$
and is probed by different observables.

In addition, \citet{Hu_GDM98} considered a ``viscosity parameter''
$c^2_{{\rm vis},\,a}(z,k)$
(defined by the ratio of the dynamical change of 
 anisotropic stress to the velocity gradient). 
However, realistically the evolution 
of anisotropic stress is determined not only by the velocity gradient 
but also by additional degrees of freedom,
e.g., by the $\ell =3$ multipole or polarization.
Therefore, in our study we will not use $c^2_{\rm vis}$ but will work directly with 
$\sigma_a(z,k)$.

A dark sector 
that couples negligibly to the other species
can be assigned a  covariantly conserved energy-momentum tensor $T^{\mu\nu}_{(a)}$
through the standard procedure of metic variation \cite[e.g.,][]{LandLifshII}.
Given the value of the sector's background energy density~$\rho_a$ 
at some moment 
(e.g. the present density, set by $\Omega_ah^2$),
$w_a(z)$~determines the density at all times from 
\be
\dot\rho=-3\Ha(1+w)\rho
\lb{dot_rho}
\ee
(the subscript~$a$ is dropped from now on.)
The evolution of density perturbation~$d$ follows by
linearly perturbing the equation of energy conservation $T^{0\nu}{}_{;\nu}=0$:
\be
\dot d + \pd_i v^i = Q
\lb{dot_d}
\ee
with\footnote{
   $\dot p/\dot\rho$, appearing in eq.\rf{Q_def},
   can be related to the background equation of state $w$ 
   as $\dot p/\dot\rho=w-\frac{\dot w}{3(1+w)\Ha}$.
}
\be
Q \equiv \fr{\dot\rho\delta p-\dot p\delta\rho}{(\rho+p)^2}
         = -3\Ha\left(c^2_{\rm eff}-\fr{\dot p}{\dot\rho}\right)\delta^{(c)}.
\lb{Q_def}
\ee
Linearization of $T^{i\nu}{}_{;\nu}=0$ 
yields the evolution of momentum-averaged velocity:\footnote{
  Eqs.\rf{dot_v}\,--\,\rf{Fs_def} are simpler
  in terms of the velocity potential: for $v^i=-\pd_i u$,
  they read 
$$
\dot u + \Ha u = \Phi + c^2_{\rm eff}\,\delta^{(c)} - \fr23\,\Nb^2\sigma.
$$
}
\be
\dot v^i + \Ha v^i = - \pd_i\Phi - \pd_i f_p + F^i_{\sigma} 
\lb{dot_v}
\ee
with
\be
f_p &\equiv& \fr{\delta p-\dot p u}{\rho+p} = c^2_{\rm eff}\,\delta^{(c)},
\lb{fp_def}\\
F^i_\sigma &\equiv& - \fr{\pd_j\Sigma^{ij}}{\rho+p} = \fr23\,\pd_i\Nb^2\sigma.  
\lb{Fs_def}
\ee
We remember that $\delta^{(c)}$ is given by eq.\rf{delta^c_Newt},
and the gravitational potentials are determined by the linearized 
Einstein equations\rf{Poiss_tradit} and\rf{Psi-Phi}.
These equations are closed after
$c^2_{{\rm eff}}$ and $\sigma$ 
are specified by the species' internal dynamics.
The expansion of perturbations over Fourier modes splits
the equations into uncoupled systems for each mode,
with its own $c^2_{{\rm eff}}(k)$ and $\sigma(k)$.

Similarly to $\sigma(k)$, the stiffness $c^2_{\rm eff}(k)$ 
is manifestly gauge-invariant but not measurable locally.

\subsection{Other parameterizations}
\label{sec_sum_param}

Any of the above sets $\{\Ha$, $\Phi$, $\Psi\}$,
$\{\rho$, $\delta^{(c)}$, $\sigma\}$, or 
$\{w$, $c^2_{\rm eff}$, $\sigma\}$
parameterizes all properties of arbitrary dark sectors
that can be constrained by any probe of background and scalar perturbations.
It may nevertheless be useful to explore other 
internal properties of the dark species.
Certain properties, e.g.\ particle masses and cross-sections, 
that are not readily extractable from the above sets
may be important for particle-physics models.
Other properties may have more distinctive 
or less degenerate observable signatures.
An example appears in Sec.~\ref{sec_speed},
discussing the dark characteristics
that control the additive shift of the CMB peaks.

Of course, yet alternative parameterizations of the internal dark dynamics 
may also be considered.
For example,\ctt{Linder_gamma05} suggested to approximately quantify
the growth of cosmic structure by a growth index parameter~$\gamma$.
This approach may be useful for utilizing LSS
to probe modified gravity or non-standard long-range interaction. 
We do not pursue it here, 
being interested in more universal descriptions
that are applicable to the CMB as well.

\section{Mapping dark dynamics to observable features}
\label{sec_impacts}

In this, main, section we identify the observable signatures of the dark properties.
Specifically, we consider the signatures of: 
anisotropic stress, the dark species' stiffness,
the propagation speed of their inhomogeneities,
and their clustering.
The observational probes considered are  
the transfer functions for cosmic structure
and the power spectra of the CMB.

\subsection{Anisotropic stress}
\label{sec_anis_stress}

Already in the standard cosmological model,
significant anisotropic stress is generated in the radiation era 
by freely streaming neutrinos.
The anisotropic stress in our universe may differ from 
the standard three-neutrino expectation
because of, for example,
additional free-streaming relativistic species, which would enhance the stress.
Or, it may differ because of non-minimal neutrino 
couplings\ct{Chacko:2003dt,Chacko:2004cz,Beacom:2004yd,Okui:2004xn,
             Hannestad:2004qu,Grossman:2005ej},
which would locally isotropize neutrino velocities
and reduce or eliminate their anisotropic stress. 
Anisotropic stress~$\sigma$ changes $\Phi$ and $\Psi$\ct{MaBert95}
and consequently the CMB gravitational driving term~$D$\rf{D_def}
even on superhorizon scales.
Thus the $\sigma$ of dark species affects the CMB modes
since the earliest stages of horizon entry.

The left panel of Fig.~\ref{fig_CMB_rad} shows  
the evolution of gravitationally coupled photon and neutrino perturbations  
in the radiation era for various assumptions about neutrino abundance and properties.
The plots are obtained by numerically integrating
the corresponding equations from \cite{B06}.
The oscillating, red curves give the CMB overdensity~$d_\gamma(\tau,k)$.
The falling, brown curves show the driving term $D(\tau,k)$, 
equal to $-3(\Phi+\Psi)$ in the radiation era.

As evident from this figure,
the photon overdensity~$d_\gamma$ in the presence of freely streaming neutrinos 
(wider solid oscillating curve)
is suppressed with respect to~$d_\gamma$ of a neutrinoless model (dashed oscillating curve).  
The suppression of the CMB perturbations by neutrinos
can be related to the reduction of $D$ early during horizon entry,
with the CMB modes then experiencing a smaller initial boost.

The suppression of $D$ and, consequently, of the CMB oscillations
in the Newtonian gauge is due to $\sigma_\nu$ being a direct source 
of gravitational potentials.
It is not due to the damping of neutrino perturbations by
the viscosity effect of $\sigma_\nu$, eqs.\rf{dot_v} and\rf{Fs_def}.
We can verify this by repeating the calculations for the standard model 
with three neutrinos but with the following modification:
We remove the gravitational effect of anisotropic stress
in eq.\rf{Psi-Phi}, i.e., set $\Phi=\Psi$.
We make no other changes in the integrated equations,
i.e., preserve the Poisson equation\rf{Poiss_tradit} 
and use the standard dynamical equations for all species' perturbations.
In particular, we retain the $\sigma_\nu$ viscosity term
in the ${\dot v}_{\!\nu}$ equation [c.f.\ eqs.~(\ref{dot_v},\,\ref{Fs_def})].
The results of this calculation, 
with the ``gravity of $\sigma_\nu$ removed''\footnote{
  The performed removal of $\sigma_\nu$ from the gravitational
  equation\rf{Psi-Phi} but not from the Euler equation\rf{dot_v}
  violates the covariance of the Einstein equations.
  Therefore, the separation of the effects of anisotropic stress
  into ``gravitational'' and ``dynamical''
  should not be considered beyond the context of the 
  Newtonian gauge. 
  We could preserve the Einstein equations by removing 
  $\sigma_\nu$ from both eqs.\rf{Psi-Phi} and\rf{dot_v}.
  The result would be the neutrinoless model,
  described  on the left panel of Fig.~\ref{fig_CMB_rad} by the dashed curves;
  again, showing no suppression of the CMB amplitude.
},
are plotted on the left panel of Fig.~\ref{fig_CMB_rad}
with the thinner solid curves.
Switching off the direct gravitational effect of $\sigma_\nu$
is seen to lift the neutrino suppression of the subhorizon CMB oscillations.

On the scales that enter the horizon in the radiation era ($\ell\gg 200$)
the suppression of the amplitude of the CMB oscillations, $A_\gamma$, 
can be calculated analytically 
in leading order in $R_\nu\equiv\rho_\nu/(\rho_\gamma+\rho_\nu)$\ct{BS04}.
For all physical values $0\le R_\nu \le1$,
this suppression is well fitted by\ct{HuSugSmall96} 
\be
\frac{A_\gamma}{A_\gamma(R_\nu\to0)} \approx
          \left(1+\fr{4}{15}\,R_\nu\right)^{-1},
\lb{Agamma_suppr}
\ee
where $R_\nu$ is varied at unchanged
fluctuations of the primordial curvature $\zeta$\ct{zeta_orig}.
This variation implies fixed inflationary physics,
although then the superhorizon values of both
$\Phi$ and $\Psi$ change with $R_\nu$.
The oscillations in the CMB power spectra 
of temperature, polarization, or their cross-correlation 
for $\ell\gg 200$ are suppressed by the square of\rf{Agamma_suppr}.

To summarize,
the suppression of the CMB oscillations by freely streaming neutrinos 
is due to neutrino anisotropic stress.
In the Newtonian gauge the suppression may be attributed 
to the anisotropic stress acting as a source of metric perturbations, 
impacting CMB photons. 
The damping of bulk velocity and density perturbations of neutrinos
by streaming plays almost no role in the suppression of CMB fluctuations.

~

~

\subsection{Stiffness $c^2_{\rm eff}$}
\label{sec_stiffness}

The quantity $c^2_{\rm eff}(z,k)\equiv \delta p^{(c)}/\delta\rho^{(c)}$ 
describes the stiffness of the dark medium to perturbations with a wavevector~$k$.
(We avoid calling $c^2_{\rm eff}$ ``the sound speed,'' 
not to mix it with $c_p$ of  Sec.~\ref{sec_speed}.)
The stiffness $c^2_{\rm eff}$ affects directly the evolution of the dark species' overdensities 
and peculiar velocities.
The direct impact of $c^2_{\rm eff}$ on the evolution of overdensity,
eq.\rf{dot_d}, is described by the term $Q$\rf{Q_def},
proportional to the so called ``non-adiabatic'' pressure 
$\delta p-(\dot p/\dot\rho)\,\delta\rho$.
The velocities are accelerated
by pressure gradient, the term 
$-\pd_i f_p=-\pd_i(c^2_{\rm eff}\delta\rho_a^{(c)})$ in eq.\rf{dot_v},
directly proportional to $c^2_{\rm eff}$.
For perfect fluids and $k\gg\Ha$, 
the quantity $c_{\rm eff}(k)$ gives the (phase) velocity of acoustic waves.
In general, however, $c^2_{\rm eff}$
need not be related to the speed of perturbation propagation,
which will be studied in Sec.~\ref{sec_speed}.

The dependence of the dark perturbations' evolution
on their stiffness $c^2_{\rm eff}$ is reflected in the perturbations'
contribution to metric inhomogeneities.
Through the latter, the dark stiffness
affects the visible species \cite[e.g.,][]{Erickson:2001bq}.
To be specific, we assume that superhorizon perturbations are adiabatic.\footnote{
  The following arguments remain valid under a considerably milder 
  condition: It is sufficient that the primordial perturbations 
  in the probed dark species are {\it internally\/} adiabatic\ct{BS04}.
  In particular, this condition is satisfied automatically 
  for any one-component fluid.
}
Then the comoving overdensity $\delta^{(c)}$, eq.\rf{delta^c_def},
vanishes in the superhorizon limit.
Consequently, $\dot d_a$ [eqs.~(\ref{dot_d}-\ref{Q_def})]
and $\dot v_a$ [eqs.~(\ref{dot_v}-\ref{fp_def})] depend on $c^2_{\rm eff}$
only at the order $O(k^2\tau^2)$.
The sourced potentials $\Psi$ and $\Phi$ therefore
become sensitive to $c^2_{\rm eff}$
only at a late period of the horizon entry, 
also only at the order $O(k^2\tau^2)$.
Note that, as discussed in Sec.~\ref{sec_anis_stress},  
anisotropic stress, in contrast, changes the potentials even in the 
superhorizon limit $k\tau\to 0$.
Thus $c^2_{\rm eff}$ affects 
the observable species at a noticeably {\it later\/}
stage of the horizon entry than~$\sigma$ does.

The lateness of the impact of $c^2_{\rm eff}$ is clearly seen
on the right panel of Fig.~\ref{fig_CMB_rad},
which shows the joint evolution of perturbations
in a photon fluid and a classical scalar field~$\phi$ (quintessence).
The quintessence background density is set to track radiation ($w_\phi=1/3$)
but its stiffness exceeds that of the photon fluid 
($c^2_{{\rm eff},\phi}=1>c^2_{{\rm eff},\gamma}=1/3$, e.g.,\ctt{Hu_GDM98}).
Either the perturbations of the quintessence (right panel)
or of freely streaming neutrinos (left panel and previous 
Sec.~\ref{sec_anis_stress}) 
decrease the CMB-driving sum $\Phi+\Psi$.
With neutrinos, 
$\Phi+\Psi$ was decreased already on superhorizon scales by
neutrino anisotropic stress ($\sigma_\nu\not=0=\sigma_{\gamma b}$).
On the other hand, in agreement with the previous arguments,
$\Phi+\Psi$ is seen to be affected by the stiff inhomogeneities
of quintessence, whose anisotropic stress is zero,
at a later time.

There are several observable consequences of the lateness 
of the influence of dark stiffness on potentials.
First, while the early reduction of $\Phi+\Psi$
by $\sigma_\nu$ suppressed the amplitude of the acoustic oscillations
in~$d_\gamma$,
the later reduction of $\Phi+\Psi$ 
by large $c^2_{{\rm eff},\phi}$
slightly {\it enhances\/} this amplitude.
The right panel of Fig.~\ref{fig_CMB_rad} explains this paradox:
The reduction of $\Phi+\Psi$ by quintessence perturbations
increases the negative driving of photon overdensity
when $\dot d_\gamma$ is on average negative.
As a result, the acoustic oscillations gain energy
and their amplitude increases.

\begin{figure*}[t]
\centerline{\sc\footnotesize \qquad CDM modes}
\centerline{\includegraphics[width=12cm]{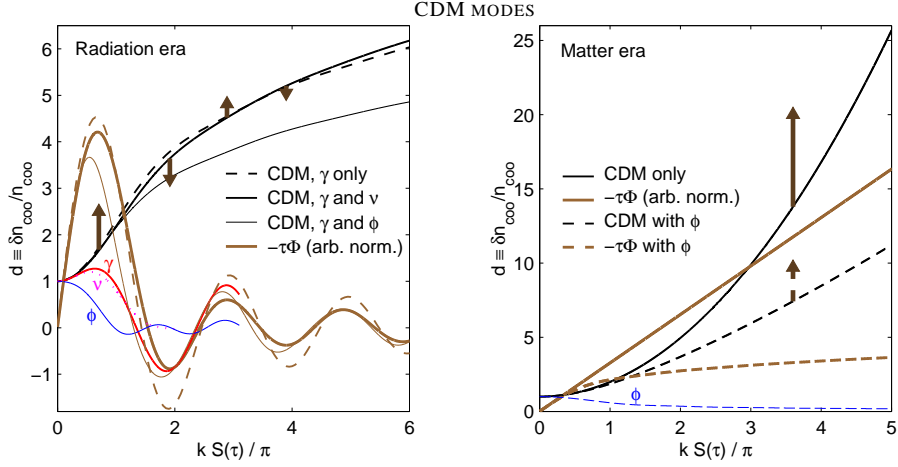}}
\caption{
Growth of CDM overdensity~$d_c(\tau,k)$ (rising black)
during radiation domination (left) and matter domination (right). 
Brown curves show the value of $\tau\Phi$,
responsible for the ultimate growth of $\delta\rho_c/\rho_c$
by a future time of its observation, eq.\rf{growth_grf}.
---\,Left: coupled photons only (dashed),
photons plus 3 standard neutrinos (wide solid),
and photons plus a tracking ($w_\phi=1/3$) scalar field~$\phi$
replacing the neutrino density of the previous model (thin solid).
---\,Right: pressureless matter only (solid),
and a model with equal densities of matter 
and a scalar field that tracks it, $w_\phi=0$ (dashed).
}
\label{fig_CDM}
\end{figure*}

The lateness of the stiffness impact has
another important consequence.
Namely, the ratio of the matter response over the CMB response
is considerably larger to probing~$c^2_{\rm eff}$ than to probing~$\sigma$.
This is evident from comparing the left panel of Fig.~\ref{fig_CDM}, 
showing CDM overdensity (black, rising curves),
and the earlier Fig.~\ref{fig_CMB_rad}, for the CMB overdensity,
all in the radiation era.
In Fig.~\ref{fig_CDM}, CDM overdensity $d_c$
is plotted for: the photon-only model, where $\sigma=0$ (dashed); 
the standard model with $59\%$ of energy density in coupled photons 
and $41\%$ in streaming neutrinos (wide solid);
and a fictitious model with the same $59\%$ of energy density in photons
but the remaining $41\%$ carried  by quintessence (thin solid).
We see that the enhanced stiffness of the dark component of the third model
impacts the matter an order of magnitude more strongly
than the anisotropic stress of the second model.
On the other hand, the CMB perturbations, 
studied on the left and right panels of Fig.~\ref{fig_CMB_rad},
are affected by neutrinos and quintessence
of the second and third models comparably.

These results have a simple explanation.
The early impact of~$\sigma$
on matter velocities is washed out over time by the 
Hubble friction [$\Ha v^i_c$ term in eq.\rf{dot_c}].
The Hubble friction,
acting on massive CDM and baryons,
damps less the $c^2_{\rm eff}$-dependent late impact 
of dark perturbations on matter.
The overdensity of relativistic photons,
on the other hand, is unchanged by Hubble friction
[see eq.\rf{dot_gk}, where $R_b$ is negligible].
Hence the present CMB anisotropy responds comparably to either of the impacts.

The relatively high ratio of 
the matter over the CMB responses to the dark stiffness
should help distinguish an excess of relic relativistic particles 
in the radiation era from
a subdominant tracking quintessence.
Both the decoupled relativistic relics and quintessence
have a similar nondegenerate phase-shift signature in the CMB power
spectra \citep[][more in Sec.~\ref{sec_speed} below]{BS04}.
This signature will soon considerably tighten the constraints
on neutrinos or an early quintessence  
with the enhanced angular resolution of 
CMB experiments.
Yet it does not discriminate between the two scenarios 
if a nonstandard signal is observed.
The discrimination can, however, be achieved by
combining the CMB data with accurate measurements of matter power,
together sensitive to the strong effect of quintessence on the ratio 
of the CMB and matter power.
The discrimination should be facilitated by the fact that
tracking quintessence and an excess of relativistic species 
change this ratio in opposite directions.

\subsection{The speed of sound or streaming}
\lb{sec_speed}

Dark inhomogeneities were seen to affect
the amplitude of the oscillations in the CMB temperature and polarization power spectra.
In addition to the  heights of the acoustic peaks, 
the CMB spectra are characterized by the peaks' positions~$\ell_n$.
The period,~$\ell_A$, of the acoustic oscillations 
is well known to be controlled entirely by
background geometry and the properties of the photon-baryon plasma.
(Namely, $\ell_A=\pi r\dec/S\dec$,
where $r\dec$ is the distance to the surface of CMB decoupling 
and $S\dec=\int_0^{\tau\dec}c_{\!\gamma b}\,d\tau$ 
is the corresponding size of the acoustic horizon.)
On the other hand, the overall additive shift of the peak's sequence,
i.e., the parameter $\Delta \ell$ in an approximate relation
$\ell_n\sim n\ell_A+\Delta \ell$ is a robust characteristic of dark perturbations.\footnote{
  The positions of individual peaks receive additional corrections
from the changes of the peaks' shape, such as due to the Silk damping and 
(for temperature) falling Doppler contribution.
These corrections are fixed by the background properties
and should not spoil the constraints on the overall shift~$\Delta \ell$. 
}

The shift of the peaks in $\ell$ is determined by the shift $\Delta\varphi$
of the temporal phase of the subhorizon modes,
$d\sg= A\sg \cos(kc_{\gamma b}\tau-\Delta\varphi)$, as 
$\Delta \ell=\ell_A\,\Delta\varphi/\pi$.
This shift of phase is tightly connected to the locality of the
inhomogeneous dynamics\ct{BS04,B05Trieste}.
This becomes manifest
in the real-space description of perturbation evolution
of \citet{BB_PRL,BB02}, using plane-parallel Green's functions. 
The Green's function of $d\sg$ in the radiation era,
when the modes forming the acoustic peaks enter the horizon, 
was found in \citet{BS04}.
Its Fourier transform yields
\be
\Delta\varphi \simeq -\pi\sqrt3\ (\Phi+\Psi)_{|x|=c_s\tau}
\ee
(for $|\Delta\varphi|\ll 1$.)
Here, the initial conditions are assumed adiabatic,
and $c_s\equiv c_{\gamma b}=1/\sqrt3$ is the speed of sound in the radiation era.
The term $(\Phi+\Psi)_{|x|=c_s\tau}$
is the value of $\Phi+\Psi$ at the acoustic horizon
of a perturbation that is initially localized in space 
to a thin plane: $d(\tau\to0,x)=\delta_D(x)$,
where $\delta_D(x)$ is the Dirac delta function.
For adiabatic initial conditions,
$(\Phi+\Psi)_{|x|=c_s\tau}$ is fully determined by
the perturbations that
propagate beyond the acoustic horizon.
In particular, $(\Phi+\Psi)_{|x|=c_s\tau}=0$
if none of the dark species support perturbations that
propagate faster than the acoustic speed~$c_s$. 
[For a proof, accounting for subtleties of inhomogeneous gravitational dynamics
on large scales, see Appendix~B of \citet{BS04}].

Two commonly considered types of cosmological species
do support perturbations that propagate faster than~$c_s$.
These are decoupled relativistic neutrinos and quintessence,
in both of which perturbations propagate at the speed of light.
The increase of the energy density in either of these species
by one effective fermionic degree of freedom
displaces the peaks of the CMB temperature and E-polarization spectra
by $\Delta l_{+1\nu}\simeq -3.4$ for neutrinos\ct{BS04} 
and $\Delta l_{+1\varphi}\simeq -11$ for tracking quintessence\ct{B06}.
This nondegenerate effect will lead to tight constraints on the 
abundance\ct{Lopez:1998aq,Bowen02,BS04,Perotto:2006rj}
and interaction\ct{Chacko:2003dt}   %Friedland:2007vv} 
of the relic neutrinos from upcoming CMB experiments.
These constraints notably improve with higher angular resolution in temperature and polarization
as new data at higher $\ell$'s allow better identification of the oscillations' phase.
[The polarization channel, where the acoustic peaks are more pronounced,
is especially useful\ct{BS04}.]

Thus a crucial characteristic that affects the acoustic phase 
is the velocity $c_p$ of the wavefront of a localized perturbation
of dark species, shifting the phase if and only if 
$c_p>c_s\approx 1/\sqrt3$.
The velocity $c_p$ of a perfect fluid or a classical field
is determined by the stiffness~$c^2_{\rm eff}$.
However, in general $c_p$ and $c_{\rm eff}$ are independent.
As a notable example, free-streaming relativistic particles
have $c_{\rm eff}(k)\equiv 1/\sqrt3$ but $c_p=1$.
Unlike $c_{\rm eff}(k)$, 
$c_p$ is defined irrespectively of any gauge and is locally measurable.
Finally, as discussed in Sec.~\ref{sec_stiffness} and the present section,
the quantities $c_{\rm eff}$ and $c_p$ map to different observable signatures. 
Thus for gaining robust knowledge of the nature and kinetics 
of dark radiation, including neutrinos and
possible early quintessence, it is important to constrain observationally
both $c_{\rm eff}(k)$ and $c_p$.

\subsection{$\Phi$ and $\Psi$; clustering in the dark sectors}
\lb{sec_clustering}

In the three previous subsections we analyzed the gravity-mediated
signatures of the dark stresses.
Now we shift our attention to the observational manifestations of 
potentials $\Phi$ and $\Psi$ themselves,
irrespective of their cause or even of the validity of the Einstein equations. 
As described quantitatively in this subsection,
the impact of the gravitational potentials on 
large-scale structure (LSS) and the CMB differs 
in several essential ways. 
Consequently, the corresponding constraints will be rather complimentary.

Before considering the observable impact of the metric potentials,
let us summarize the scenarios in which 
the potentials are expected to differ from the  
predictions of the standard cosmological model.
As compared to a typical model 
with CDM and quintessence dark energy  
(including $\Lambda$CDM model as a limiting case),
different $\Phi$ and $\Psi$ 
in a given background $w(z)$
are expected:
\begin{enumerate}
\item[(a)]
During horizon entry for practically
any other dark energy model.
\item[(b)]
On any scales for models with modified gravity.
\item[(c)]
On subhorizon scales for models 
with non-canonical kinetic term for the field\ct{ArmendarizMukhanov_kEss00},
models with warm dark matter\ct{Blumenthal:1982mv,Olive:1981ak} 
or with interacting dark matter\ct{Carlson_et_al_92,deLaix:1995vi,Spergel:1999mh}.
More mundane physics affecting the subhorizon potentials
includes thermal, radiative, or magnetic pressure on baryons,
and various astrophysical feedback mechanisms (supernovae, central black holes, etc.)
\end{enumerate}

\subsubsection{Matter response}

Cosmic structure, which growth in the matter era is driven by potential~$\Phi$, 
is very sensitive to modifications of~$\Phi$ by new physics at low redshifts.
As noted in Sec.~\ref{sec_stiffness},
the Hubble friction diminishes its sensitivity toward higher redshifts.
%The self-gravity of subhorizon inhomogeneities is
%the primary mechanism that drives active structure growth in the matter era.\footnote{
%   In contrast, the weak subhorizon evolution of matter perturbations in the radiation era
%   is predetermined by the gravitational boost of the species' velocities at the horizon entry.
%}
The matter power spectrum and other characteristics of 
cosmic structure are therefore the natural probes of the scenarios 
in the above classes (b) and~(c).

An example is shown on the right panel of Fig.~\ref{fig_CDM}.
For both scenarios displayed on this panel the background
is Einstein--de Sitter.
The solid curves show the growing matter overdensity
and gravitational potential in a pure CDM phase.
[The plotted potential is weighted by $\tau$ according to eq.\rf{growth_grf}.]
The dashed curves display the same quantities
for a toy scenario in which CDM constitutes only $50\%$ of the energy density,
while the other $50\%$ is in a classical scalar field
with $w_\phi=w_{\rm CDM}\equiv0$.
Since for the field $\delta p_\phi^{(c)}/\delta\rho_\phi^{(c)} = c^2_{\rm eff}=1$,
its clustering is suppressed and its perturbations do not
contribute to the subhorizon potential.
Correspondingly, the structure growth in the second scenario is seen to be significantly suppressed.
Note that both scenarios have identical background expansion
and the standard laws of gravity.

Quantitatively, the impact of the potential $\Phi$ over time $(\tau,\tau+\Delta\tau)$
on matter overdensity $d_c$ 
observed at a later time $\nbrk{\tau'\gg\tau}$ scales roughly as $(\tau\Delta\tau)\Phi$.
The prefactor~$\tau\Delta\tau$ quantifies the stronger suppression of
an early impact by the higher Hubble friction,
experienced by matter particles during the faster cosmological expansion.
The derivation follows straightforwardly from eq.\rf{d_c_sol}:
A nonzero potential $\Phi(\tau)$ over the time interval 
$(\tau,\tau+\Delta\tau)$ contributes to the matter overdensity 
at a later time  $\tau'$
as\footnote{
   The last estimate in eq.\rf{growth_grf} assumes that $w\le1/3$,
  so that $a(\tau)$ grows as $\tau$ or faster. 
  The zero of conformal time $\tau$
  is chosen, as usual, as $\tau\to0$ when $a\to0$.
  Logarithmic corrections that appear for $w=1/3$
  (radiation domination) are ignored.
}
\be
\Delta\!\lf( \fr{\delta\rho_c}{\rho_c} \rt)_{\!\!\tau'} = \Delta\, d_c(\tau')
        &=& - k^2\,a(\tau)\Phi(\tau)\,\Delta\tau \int_{\tau}^{\tau'} \frac{d\tau''}{a(\tau'')}\nonumber\\
        &\sim& - k^2\Phi(\tau)\,\tau\,\Delta\tau. \quad
\lb{growth_grf}
\ee

In the models with standard gravity, CDM,
and {\it insignificant clustering of dark energy\/},
the dark energy perturbations
can leave their mark on the potential only during horizon entry.
By eq.\rf{growth_grf}, the corresponding early potential
contributes to structure growth much less than 
the potential generated long after the entry by the clustered CDM and baryons.
Thus in the absence of the non-standard physics of the types (b)\ or~(c) above,
it is justified to consider an approximate consistency relation 
between the background equation of state~$w(z)$
and the growth of cosmic 
structure\ct{KnoxSongTyson05,Ishak:2005zs,Chiba:2007rb}.

As we see next, the situation is essentially {\it opposite\/} 
when the dark sectors are probed by the primary anisotropies of the CMB.

\subsubsection{CMB response}

Metric perturbations have a dramatic impact on 
CMB temperature anisotropies. 
In sharp contrast to large scale structure,
primary CMB anisotropies (generated by linear evolution)
depend most strongly on the values and evolution of gravitational potentials
{\it during horizon entry\/}. 
They are only mildly sensitive to the potentials on subhorizon scales.
(This need not to apply to secondary anisotropies.)

The scalar perturbations of the CMB respond primarily to the sum
of the Newtonian-gauge potentials $\Phi+\Psi$ on all scales
and all times (Sec.~\ref{sec_CMB}). 
While the inertial drag of the CMB by baryons, affected by $\Phi$ alone,
is important for the CMB sensitivity to the baryon density 
and is noticeable around decoupling\ct{HuSugSmall96}, 
it never dominates the evolution of CMB perturbations.
For developing a general understanding of the CMB sensitivity 
to dark dynamics we will often ignore it.

We stressed in Sec.~\ref{sec_uniq} that
dark inhomogeneities at a (conformal) time~$\tau$ 
and the metric perturbations generated by them 
affect only the modes with $k\gtrsim \Ha\sim 1/\tau$.
The corresponding dynamics of CMB perturbations differs qualitatively
in a tightly coupled regime, 
prior to hydrogen recombination at $z\rec\sim 1100$,
and in a streaming regime, after recombination.
We will consider the CMB response to the metric perturbation $\Phi+\Psi$
at these two epochs in turns.
In the subsequent Sec.~\ref{sec_cl_resp_quant}
we will quantify this response by simple analytical formulas.
We will see that despite numerous differences in the evolution of CMB modes
before and after recombination, the CMB response to $\Phi+\Psi$
at the respective epochs is very similar.

\begin{figure}[t]
\centerline{\includegraphics[width=7cm]{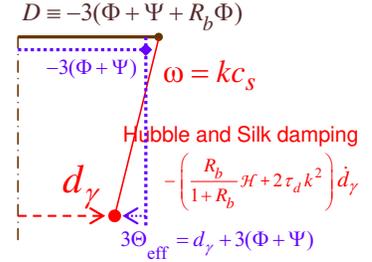}}
\caption{An accurate mechanical analogy for general-relativistic evolution\rf{dot_gk}
of an acoustic CMB mode.
The CMB overdensity~$d\sg$ equals
the denoted distance to the tip of a pendulum with an internal
frequency $\omega=kc_s$ and its suspension point 
driven as specified by the gravitational driving term~$D$, eq.\rf{D_def}.
For adiabatic initial conditions the evolution
starts with $d_{\rm in}=3\zeta_{\rm in}$,
where $\zeta_{\rm in}$ is the superhorizon value of the Bardeen curvature.
In the radiation era, initially,
$D_{\rm in}=\frac43(1+\frac15R_\nu)/(1+\frac4{15}R_\nu)d_{\rm in}$
and in the matter era $D_{\rm in}=\frac65d_{\rm in}$\ct{BS04}, 
c.f.\ Figs.~\ref{fig_CMB_rad} and~\ref{fig_CMB_mat}.
The advantages of this approach over 
considering $\delta T^{(\rm Newt)}\!/T$
or $\Theta_{\rm eff}=\delta T^{(\rm Newt)}\!/T+\Phi=\fr13d\sg+\Phi+\Psi$
as independent dynamical variables include:
(A)~Independence of the gravitational force that drives $\ddot d\sg$ 
   from $\ddot d\sg$ itself.
(B)~Epoch- and scale-independence of the relation between 
   the superhorizon perturbation~$d_{\rm in}$ 
   and the inflation-generated conserved curvature~$\zeta_{\rm in}$
   ($d_{\rm in}=3\zeta_{\rm in}$).
(C)~Direct cause-effect connection between local physical dynamics
   and the apparent changes of the variable that describes CMB perturbations. 
}
\label{fig_CMB_analogy}
\end{figure}

~

\subsubsubsection{Coupled regime ($z\gtrsim z\rec \sim 1100$)}

Any CMB mode prior to recombination
evolves according to eq.\rf{dot_gk} of a driven damped harmonic oscillator.
Fig.~\ref{fig_CMB_analogy} presents an equivalent mechanical 
system---a pendulum whose evolution is described by the same equation.
The pendulum's internal frequency 
is $kc_s$ and its pivot is moved by a distance 
\be
D=-3(\Phi+\Psi+R_b\Phi).
\ee
The CMB overdensity $d\sg(\tau)$ then numerically equals
the distance to the pendulum's tip, as marked on the figure. 
To account for the Hubble and Silk damping,
the pendulum may be imagined submerged in a viscous fluid
with appropriate time-dependent viscosity.
All the influence of the metric perturbations on $d\sg$ 
is transferred through the
(generally, time-dependent) position of the pendulum's pivot~$D(\tau)$. 

\begin{figure*}[t]
\centerline{\sc\footnotesize CMB, Matter era \qquad}
\centerline{\includegraphics[width=13cm]{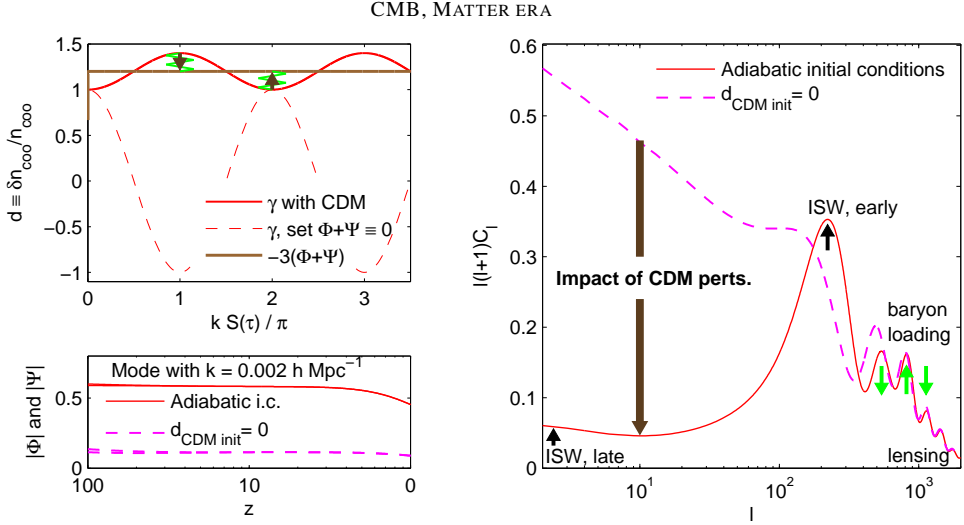}}
\caption{ Gravitational suppression of CMB temperature anisotropy by 
growing cosmic structure in the matter era 
(5-fold for $\Delta T/T$, 25-fold for power $C_l$). 
---\,Left top: Evolution of~$d\sg$ and the driving term~$D$ 
during matter domination with $R_b$ set negligible (solid).
Evolution of~$d\sg$ from the same primordial perturbation
if the metric becomes homogeneous before the horizon entry (dashed).
---\,Right: Suppression of the CMB temperature power~$C_l$ for $\ell\lesssim 100$
by CDM inhomogeneities.
The solid curve shows~$C_l$ in the concordance $\Lambda$CDM 
model with  adiabatic initial conditions.
The dashed curve describes the same model 
with changed initial CDM perturbations:
$d_{\rm CDM}$ is artificially set to zero on superhorizon scales
(the superhorizon $d\sg$, $d\snu$, and $d_b$ are unchanged).
Then CDM inhomogeneities and the associated potential in the matter era
are reduced.
The smoother metric suppresses the CMB power at $\ell\lesssim 100$ less,
as the  $C_l$ plots (obtained with {\sc CMBFAST}) clearly show.
---\,Left bottom: The plots of $\Phi$ and $\Psi$ transfer functions confirm that the last model
has smaller $\dot\Phi+\dot\Psi$,
hence, the enhancement of its power at low $\ell$ 
cannot be attributed to the ISW effect.}
\label{fig_CMB_mat}
\end{figure*}

This mechanical analogy of the acoustic CMB evolution
makes intuitive the following important facts:
Foremost, the CMB overdensity is influenced strongly by
the {\it values\/} of the potentials at horizon entry.
Also, as already observed in Sec.~\ref{sec_stiffness},
the sensitivity and even the sign of the CMB response
to the subsequent changes of the potentials depend strongly
on the {\it time\/} and {\it duration\/} of the change.

When at horizon entry the driving term~$D$ is close 
to the superhorizon value of $d\sg$ 
(as it usually is for adiabatic perturbations)
and when $D$ {\it does not decay quickly\/} during the entry
then CMB temperature anisotropy is suppressed dramatically.
For a specific example, roughly reflecting the evolution 
between radiation-matter equality and decoupling
($3\times 10^3 > z > 1\times 10^3$),
let us ignore the baryon-photon ratio $R_b$
by taking $D\simeq -3(\Phi+\Psi)$. 
Let us evaluate the potentials in a CDM-dominated limit, 
in which $\Phi(\tau)=\Psi(\tau)=\const=-\frac15d_{\rm in}$,
where for adiabatic perturbations $d_{\rm in}\equiv d_a(\tau{\to}0)=3\zeta(\tau{\to}0)$
for all species $a$.\footnote{\lb{note_Psi15d}
 The CDM-domination result $\Phi=\Psi=-\frac15d_{\rm in}$
 follows straightforwardly 
 from eqs.~(\ref{d_c_sol},\,\ref{Poiss_my}-\ref{Poiss_my_notat}) 
 and from $\Phi=\Psi$ by eq.\rf{Psi-Phi}.
 It is also easy to derive from the known relation 
 between potentials and the traditional proper overdensity
 by remembering that for pressureless matter $d=\delta\rho/\rho-3\Psi$.
}
In these approximations, 
\be
D(\tau)=\frac65\,d_{\rm in},
\ee
being time-independent during and after the mode entry.
Then the amplitude of acoustic CMB oscillations 
is {\it suppressed 5-fold\/}\ct{B06},
which is evident from
Fig.~\ref{fig_CMB_analogy} or the left top panel of Fig.~\ref{fig_CMB_mat}.

After the entry,
a slow variation of~$D$ over a characteristic time $\delta\tau$
that exceeds the period of $d\sg$ oscillations ($\delta\tau\, k \gg1$)
has minor impact on the oscillation amplitude and phase.
In particular, this impact 
vanishes in the adiabatic limit, $\delta\tau\, k \to \infty$.
A typical temporal scale of linear variations of $\Phi$ and $\Psi$, and so of $D$,
is $\delta\tau\sim\Ha^{-1}$. 
This applies even to subhorizon scales,
where only the perturbations with negligible pressure,
hence no rapid internal oscillations, contribute to the potentials.
Thus deeply {\it subhorizon\/} CMB modes (for which $\delta\tau\, k\sim k/\Ha \gg1$)
are {\it insensitive\/} to such variations.
Conversely, the modes that are closer to horizon entry
are affected more strongly.

\subsubsubsection{Streaming regime ($z\lesssim z\rec \sim 1100$)}

Analogous results exist for the sensitivity of the CMB 
to gravitational potentials after recombination.

When the universe is matter-dominated and baryons have decoupled 
from the CMB ($10^3 > z \gg 1$) 
then $\Phi$ and $\Psi$ are equal and time-independent on all scales
that are sufficiently large to evolve linearly
and to be unaffected by the residual baryonic pressure.
For time-independent $\Phi+\Psi$ and 
negligible collision term $C_T$ in eq.\rf{dot_dig_C},
an ``effective'' CMB intensity perturbation
\be
\di\eff\equiv \di + 4(\Phi+\Psi)\qquad
\lb{di_eff_def}
\ee
is constant along the line of sight \citep{KS86,HuSug_ISW94,HuSug_toward94}.
Under a general evolution of the potentials, eq.\rf{dot_dig_C} gives
\be
\dot\di\eff + n_i\Nbi\di\eff = 4(\dot\Phi+\dot\Psi) + C_T.
\lb{dot_dieff_C}
\ee

Prior to horizon entry, the intensity perturbation~$\di$ 
is frozen and equals
$\di_{\rm in}=\fr43d_{\gamma,\,\rm in}$ \cite[see][]{B06}. 
As discussed earlier, for the modes that enter
during the domination of pressureless matter,  
$\Phi(\tau)=\Psi(\tau)=-\frac15d_{\rm in}$.
Then, for adiabatic perturbations for which $d_{\gamma,\,\rm in}=d_{\rm in}$, 
eq.\rf{di_eff_def} gives
\be
\di\eff=-\fr15\di_{\rm in}.
\ee

Similarly to the CMB evolution in the coupled regime,  
for most of these modes 
the late ISW effect due to the {\it slow\/} decay of the potentials 
at $z\lesssim 1$ does not noticeably
change $\di\eff$.\footnote{
 The solution of the transport equation\rf{dot_dieff_C}
 for a perturbation mode $\di\eff=A(\tau)\,e^{i\bk\cdot(\bx-\bn\tau)}$
 in a potential $\Phi+\Psi=F(\tau)\,e^{i\bk\cdot\bx}$ is 
\be
A(\tau)=A_{\rm in}+4\int_0^{\tau}\dot F(\tau')\,e^{i\bk\cdot\bn\,\tau'}d\tau'.
\ee
 Due to the oscillating factor $e^{i\bk\cdot\bn\,\tau'}$,
 any slow variation of the subhorizon potential over a temporal scale
 $\delta \tau$ does not affect any modes 
 except for those few with $|\bk\cdot\bn|\,\delta\tau \lesssim 1$,
 i.e., those whose wavevector is nearly orthogonal to 
 the direction of light propagation.
}
Thus, first, 
the presently observed temperature anisotropy in the considered modes
is suppressed by the potentials of the clustering matter
5-fold\footnote{
  This suppression, by a factor of 5,
  is larger than 
  an apparent factor of 2 suppression
  that is usually inferred, erroneously, from the traditional 
  description of the Sachs-Wolfe effect in terms of the 
  proper Newtonian-gauge perturbations. 
  (Indeed, for the latter 
  $\lf.\Delta T/T\rt|_{\rm observed}\simeq 
   -\frac12 (\delta T^{\rm (Newt)}/T)_{\rm decoupling}$, 
   assuming matter domination at decoupling.
   Yet this result is specific to the Newtonian gauge condition
   for the decoupling surface, on which the considered perturbations are
   superhorizon. It should not be given its traditional physical interpretation directly.) 
   See\ctt{B06} for additional discussions.
}---suppressed with respect to a fictitious scenario 
in which in the matter era $\Phi=\Psi=0$, consequently,
$\di\eff=\di_{\rm in}$. 
Second, after decoupling as well, the primary CMB anisotropies 
have little sensitivity to the potentials' evolution on subhorizon scales.

The 5-fold suppression is a physical effect,
independent of the choice of variables or gauge. It is absent
in models where the matter-era potentials decay early during horizon entry
or where the laws of gravity are modified so that matter inhomogeneities 
do not perturb the metric.
Note that if all modes that contribute to  
the CMB temperature autocorrelation $C_\ell$ at some $\ell$
are suppressed 5-fold then
the $C_\ell$ is suppressed by $5^2=25$ times.

The right panel of Fig.~\ref{fig_CMB_mat} presents
convincing evidence for the reality of the order-of-magnitude
suppression of the CMB temperature power spectrum at low $\ell$'s 
by dark matter clustering.
The plot shows the CMB temperature power spectra~$C_l$
obtained with a modified {\sc cmbfast} code\ct{CMBFAST96} for two models
both of which have the identical matter content of the concordance $\Lambda$CDM 
model\ct{WMAP3Spergel,SelSlosMcD06,TegmLRG06}.
The models differ only by the initial conditions for CDM perturbations.
In one of the models (solid curve) all species
are initially perturbed adiabatically.
In the other model (dashed curve) the CDM density perturbation 
$d_c$ is artificially set to zero on superhorizon scales
while the initial values for $d\sg$, $d\snu$, and $d_b$ are unchanged.
As a consequence, in the second model
the metric inhomogeneities in the matter era are reduced.
While this model has a smaller ISW effect 
(see the left bottom panel of Fig.~\ref{fig_CMB_mat}),
due to the smoother metric, its CMB power for $\ell\lesssim 100$
is considerably larger.

The suppression of the CMB anisotropy on large scales, 
which enter during matter domination and active growth of structure, 
should not be partly traded 
for the ``resonant self-gravitational driving'' of small-scale modes
that enter in the radiation era.
The acoustic modes entering during radiation domination 
are often said to be resonantly driven by a specially timed 
decay of their self-generated potential.
In reality, the suppression of large relative to small scales
is not very sensitive to the timing of the potential decay in the radiation era\ct{B06}.
Moreover, the same (when accounting for neutrinos, even somewhat larger)
suppression of large relatively to small scales would be observed if 
in the radiation era the metric were unperturbed, hence, 
the small-scale modes objectively could not be driven gravitationally\ct{B06}.

\subsubsection{Quantifying the response of the CMB power spectrum}
\lb{sec_cl_resp_quant}

The CMB sensitivity to gravitational potentials at various epochs
can be quantified as follows.
The temperature anisotropy of CMB radiation observed in a direction $\bn$
is given by the following integral along the line of sight \citep{CMBFAST96}:
\be
\fr{\Delta T(\bn)}{T}= \int_0^{\tau_0}d\tau\ S(\tau,\bn r(\tau)).
\lb{los_int}
\ee
Here, $\tau_0$ is time at present,
$r(\tau)=\tau_0-\tau$ (in flat models) is the radial distance along the line of sight,
and the source equals 
\be
S=\dot g \lf(\fr13d\sg+\Phi+\Psi-v_b^i n_i+Q^{ij}n_in_j\rt)+
  g\lf(\dot\Phi+\dot\Psi\rt),~~~
\lb{los_source}
\ee
with
$v_b^i$ being the baryon velocity, 
$Q^{ij}$ being determined by the radiation quadrupole and polarization, 
and a visibility function $g(\tau)=\exp\,(-\int^{\tau_0}_{\tau}d\tau/\tau_T)$
giving the probability of CMB photons to reach us unscattered.
Regardless of the random realization of the primordial curvature perturbation~$\zeta_{\rm in}$,
the physics behind the subsequently developed perturbations of potentials, densities, etc.\ 
in linear theory can be described by transfer functions
\be
T_\Phi(k,\tau)\equiv \Phi(k,\tau)/\zeta_{\rm in}(k),
\ee
with analogous definitions for other scalar perturbations:
$\Psi$, $d\sg$, velocity potential $u_b$ (s.t.\ $v_b^i=-\pd_i u_b$), etc.

The CMB constraints on the transfer functions are
derived from observational estimates of CMB angular power spectra~$C_l$.
In particular, the temperature power spectrum equals\ct{CMBFAST96}
\be
C_\ell=4\pi \int_0^{\infty}{dk\over k}\, \Delta^2_\zeta(k)\,
                         \lf|\int_0^{\tau_0}\,d\tau\,T_S(\tau,k)\,j_l(kr(\tau))\rt|^2,~~~
\lb{C_l}
\ee
where $\Delta^2_\zeta(k)\equiv k^{3}P_\zeta(k)/(2\pi^2)$
is the dimensionless primordial power of $\zeta_{\rm in}(k)$,
and $j_l$ is the spherical Bessel function.
In eq.\rf{C_l}, the full transfer function $T_S$ of the source $S$,
eq.\rf{los_source}, 
is a differential operator, with derivatives corresponding to 
the direction-dependent terms of~$S$.
In the following discussion  
we will mostly be concerned with its scalar terms,
as only they involve the gravitational potentials directly. 

It is useful to keep in mind that the dominant contribution to $C_\ell$
at a given $\ell$ comes from the modes with $k\sim \ell/r$.
Indeed, $j_l(kr)$ vanishes exponentially at smaller values of~$k$
and as $1/k$ at higher values.

\subsubsubsection{Radiation era  ($z>z\eq\sim 3000$, probed by $\ell>200$)}

By the arguments of Sec.~\ref{sec_uniq}, 
only the modes that enter the horizon before radiation-matter equality
provide information about the metric and dark dynamics in the radiation era.
Then after horizon entry
$\delta\rho/\rho\approx \delta\rho\sg/\rho\sg$ oscillates with nearly constant amplitude.
Consequently, the induced $\Phi+\Psi$ decays as $\tau^{-2}\propto a^{-2}$
[eq.\rf{Pois_subhor}].
Although the potential decay is eventually halted by the growth of structure  
since matter domination, $\Phi+\Psi$ for these modes remains
a subdominant source of CMB anisotropy in eq.\rf{los_source},
exceeded by the intrinsic photon-baryon perturbations.\footnote{
  Photon-baryon perturbations also decay on small scales due to Silk damping.  
  Estimates\ct{HuSugSmall96,Weinberg:2002kg} show that 
  the exponential Silk damping overcomes the quadratic decay of the potential 
  only at $\ell\gtrsim 4000$,
  where secondary anisotropies are expected to dominate the 
  considered primary signal.
}

Thus the impact of gravitational potentials in the radiation era
is confined to horizon entry. 
It is fully encoded in the amplitude and phase of the subsequent acoustic oscillations.
Specific cases of such an impact were considered in 
Secs.~\ref{sec_anis_stress} and~\ref{sec_stiffness}.

\subsubsubsection{From equality to recombination 
                  ($z\rec<z<z\eq$, $100<\ell<200$)}

The CMB modes that enter during and after radiation-matter equality
($z\eq\sim 3000$) oscillate in a significant gravitational potential
of growing matter inhomogeneities.
Close to $z\eq$,
the driving potential $\Phi+\Psi$ still evolves appreciably.
The potential continues to decay because of 
the residual radiation density.
It decays until recombination ($z\rec\approx 1100$) also due to
coupling of the baryonic component of matter to the CMB,
slowing the structure growth by excluding the baryons from it
\citep[e.g.,][]{HuSugSmall96}.

Altogether, these effects lead to complicated evolution of the 
modes that enter during $z\rec<z<z\eq$.
Then both the intrinsic and gravitational terms in the source of $\Delta T/T$\rf{los_source}
contribute noticeably.
As highlighted by many studies, 
the term $\dot\Phi+\dot\Psi$ in eq.\rf{los_source}
due to the continuing decay of potentials then is also large
and boosts the height of the first acoustic peak (early ISW effect).

\subsubsubsection{After recombination ($z<z\rec\simeq 1100$, probed by $\ell<100$)}

The evolution of linear perturbations becomes simple again when $z\ll z\rec$.
On the scales that enter at this epoch but before dark energy
becomes dynamically relevant, we expect that during linear evolution
$\Phi(\tau)=\Phi(\tau)=-\fr15d_{\rm in}=\const$ (footnote~\ref{note_Psi15d}).
The presently observed perturbation of CMB intensity
for almost all these modes is suppressed 5-fold. 
Our next goals will be,
first, to confirm that the contribution of these modes to $C_\ell$
is indeed gravitationally reduced by a factor of $5^2=25$.
Second, to find the sensitivity of $C_\ell$ to different values of potentials 
due to non-standard physics.
And third, to quantify the $C_\ell$ response to the evolution of potentials on subhorizon scales. 

\begin{figure}[t]
\centerline{\footnotesize $p_\ell(x_1,x_2)\equiv j_\ell(x_1)\,j_\ell(x_2)$}
\centerline{\includegraphics[width=6cm]{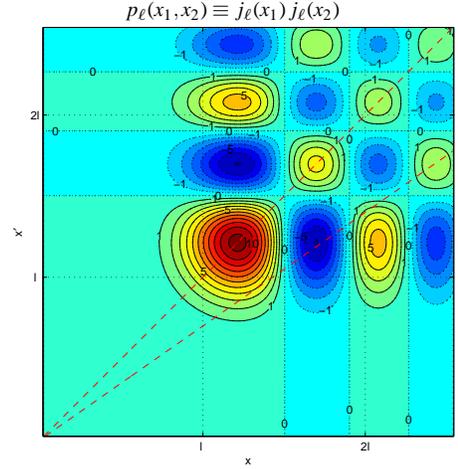}}
\caption{The isocontours of the kernel
 $p_\ell(x_1,x_2)\equiv j_\ell(x_1)\,j_\ell(x_2)$
 of projecting perturbation $k$-modes to $C_l$, eqs.~(\ref{C_l_expand},\,\ref{pl_def}).
 The figure is for $\ell=10$.
 The labels on the contours show the values of $p_{10}/10^{-3}$.
 Note that $p_\ell(x_1,x_2)\approx 0$ whenever either $x_1<\ell$ or $x_2<\ell$.
 Also note that $p_\ell\ge0$ along the line $x_1=x_2$ (diagonal dashed line),
 but $p_\ell$ oscillates through positive and negative values along 
 any line $x_1=cx_2$ with $c\not=1$ 
 (e.g., lower dashed line).
}
\label{fig_Cj1j2}
\end{figure}

We start from noting that not only do the modes of different~$k$'s
contribute to the correlation function $C_\ell$\rf{C_l} incoherently,
but coherence is also lost for same-$k$ contributions  
at widely separated times, $|\tau-\tau'|\,k\gg 1$.
To see this explicitly, we rewrite eq.\rf{C_l} as
\be
C_\ell=4\pi\!\!\int\!{dk\over k}\, \Delta^2_\zeta(k)
              \int\!\!\! d\tau\, T_S(\tau,k) 
              \int\!\!\! d\tau'T_S^*(\tau'\!,k)\,\,p_\ell(kr,kr').~~~~~
\lb{C_l_expand}
\ee
The last factor $p_\ell(kr(\tau),kr(\tau'))$ 
describes the kernel of projecting the harmonic plane-wave modes on 
a spherical multipole~$\ell$
\be
p_\ell(x,x')\equiv j_\ell(x)j_\ell(x').
\lb{pl_def}
\ee
This function for $\ell=10$, as an example, is shown in Fig.~\ref{fig_Cj1j2}.

In the integrand of eq.\rf{C_l_expand}, 
except for very low redshifts, 
$p_\ell(kr,kr')$ varies with $k$ much more rapidly than 
$T_S(\tau,k)$ and $T_S^*(\tau'\!,k)$.\footnote{
 The characteristic scales of variation are
 $\Delta k \sim \min(r^{-1},r'^{-1})$ for $p_\ell$,
 versus $\Delta k \sim \tau^{-1}$ and $\tau'^{-1}$ 
 for $T_S$ and $T_S^*$ respectively.
}
With $T_S$, $T_S^*$, and $\Delta^2_\zeta(k)$
almost unchanged over many $k$ periods of
the $p_\ell$ oscillations, 
we can see from Fig.~\ref{fig_Cj1j2} that 
positive and negative contributions to 
$\int\!(dk/k)\,p_\ell(kr,kr')$ mutually cancel whenever 
\be
|\tau-\tau'|= |r-r'|\gg 1/({\rm contributing}\ k).
\ee
Thus the contributions to $C_l$ from sources at this temporal separation
are incoherent.
In particular, we can ignore the coherence of, 
and study independently, the contributions
to $C_l$ at the horizon entry ($\tau\lesssim 1/k$)
and during subhorizon evolution ($\tau\gg 1/k$).

First, we consider a time interval 
from $\tau=0$ to $\tau_{\rm ent}\sim 1/k$.
For a mode that enters at any time after recombination,
the source\rf{los_source} forms a complete derivative:\footnote{
  For such modes, $\dot g$ significantly deviates from zero 
  only when the mode is superhorizon. 
  Then the factor that multiplies $\dot g$ in the full source\rf{los_source} 
  reduces to $(\fr13d_{\gamma,\,\rm in}+\Phi+\Psi)$,
  giving eq.\rf{los_source_late_ent}.
}
\be
S\approx {\pd\over\pd\tau}\lf[g \lf(\fr13d_{\gamma,\,\rm in}+\Phi+\Psi\rt)\rt].~~~
\lb{los_source_late_ent}
\ee
Starting from eq.\rf{C_l}, neglecting the change of $j_l(kr(\tau))$ 
over the considered time interval $(0,\tau_{\rm ent})$,
and trivially integrating the complete derivatives\rf{los_source_late_ent} 
over $d\tau$ and $d\tau'$, we obtain
\be
\delta C_\ell^{\rm (postrec~entry)}\!\!\!\approx
        4\pi\!\! \int\!{dk\over k}\, \Delta^2_\zeta(k)\,j_l^2(kr)
        \lf|g \lf(\!\fr13d_{\gamma,\,\rm in}+\Phi+\Psi\rt)\rt|^2_{\rm entry}
        \!\!\!\!\!\!\!\!\!\!\!\!\!\!\!~~~~~~
\lb{Cl_late_ent}
\ee
where the variables $d_{\gamma,\,\rm in}$,
$\Phi$, and $\Psi$ now stand for the corresponding transfer functions,
normalized by the condition $\zeta(\tau\to0,\, k)\equiv1$.
When the primordial power is nearly scale invariant,
we can neglect the $k$-dependence
of $\Delta^2_\zeta(k)$
over the range of $k\sim \ell/r$ that contribute to the integral\rf{Cl_late_ent}.
We can also ignore the variation of the transfer functions in the last parentheses
over this range.
Integration of the remaining $k$-dependent terms
by the standard formula $\int{dx\over x}\,j_l^2(x)=[2\ell(\ell+1)]^{-1}$
gives
\be
\delta C_\ell^{\rm (postrec~entry)}\!\!\!\approx
        \fr{2\pi\,\Delta^2_\zeta(\ell/r_{\rm ent})}{\ell(\ell+1)}\,
        \lf|g \lf(\!\fr13d_{\gamma,\,\rm in}+\Phi+\Psi\rt)\rt|^2_{\rm entry}
        \!\!\!\!\!\!\!\!\!\!\!\!.~~~~~~
\lb{Cl_late_ent_inv}
\ee

Eqs.\rf{Cl_late_ent} or\rf{Cl_late_ent_inv}
quantify the response of $C_l$ to the gravitational potentials at horizon entry
after recombination.
These equations confirm that
for adiabatic perturbations that enter in the matter era
the potentials $\Phi=\Psi=-\fr15 d_{\rm in}$ suppress 
the corresponding contribution to $C_l$ 25-fold.

We now consider the response of $C_l$
to changes of potentials on subhorizon scales.
Then the CMB sensitivity
can be quantified in the Limber approximation\ct{Limber54},
applied to the CMB by 
\citet{Kaiser84,Kaiser92} and \citet{HuWhite_weak_coupling95}.
With the details of the calculation given in Appendix~\ref{sec_appx},
the response of the power spectrum of CMB temperature 
to subhorizon changes of potentials is roughly
\be
\delta C_\ell^{\rm (postrec~subhor.)} 
                   \sim \fr{2\pi^2\Delta^2_\zeta(\ell/r)}{\ell^2}
                   [\delta(\Phi+\Psi)]^2\ \fr{r}{\ell\delta\tau}.\quad
\lb{Cl_resp_late}
\ee
Here $r$ is the comoving radial distance to the changing potentials 
and $\delta\tau$ is the time taken by a change $\delta(\Phi+\Psi)$,
assumed gradual.

Even for large changes of potentials ($|\delta(\Phi+\Psi)|\sim |\Phi+\Psi|$)
the observed response $\delta C_\ell^{\rm (subhor.)}$\rf{Cl_resp_late}
is suppressed relative to $C_\ell^{\rm (entry)}$\rf{Cl_late_ent_inv} 
by a factor $\fr{r}{\ell\delta\tau}$.
We argued previously that under linear evolution
generally $\delta\tau\sim \Ha^{-1}$.
Then for subhorizon scales ($r/\ell\ll \Ha^{-1}$)
the factor $\fr{r}{\ell\delta\tau}$ is much smaller than unity.
This suppression is easily understood as
the cancellation of positive and negative contributions to $\Delta T/T$
from the peaks and troughs of all perturbation modes 
except for those whose $\bk$ is almost orthogonal  
to the line of sight\ct{HuWhite_weak_coupling95}.

In addition, dark dynamics influences the CMB indirectly through 
the gravitational impact on baryons, coupled to the CMB by Thompson scattering. 
After recombination, 
the baryon-mediated linear contribution on subhorizon scales
is suppressed even more than the late ISW contribution\rf{Cl_resp_late}.
Indeed, the direct contribution of baryon velocity $v_b^i n_i$ to 
the source of CMB temperature perturbation\rf{los_source}
in the Limber limit vanishes\ct{HuWhite_weak_coupling95}:
In this limit, only the modes with $\bk$ orthogonal to $\bn$
would contribute to $C_\ell$, yet
for their scalar perturbations (with $v^i$ parallel to $\bk$)
$v_b^i n_i= 0$.
The indirect effect of weakly coupled baryonic perturbations through affecting
the photon overdensity or other photon multipoles 
in the source\rf{los_source} is also diminished relative
to the already suppressed ISW term 
by an additional factor $\delta\tau/\tau_T$, 
where $\tau_T$ is the photon free-flight time.
This factor is small for any temporal interval after CMB decoupling.

The CMB is also affected through various non-linear effects,
e.g., its lensing by cosmic structure or Sunyaev-Zeldovich effect.
Unlike the primary anisotropies, these non-linear features
may be useful probes of the dark dynamics on small scales
whenever they can be separated from the dominant primary signal
\citep[e.g.,][]{Seljak:1998nu,HuOkamoto_LensingReconstr01}.
\begin{figure}[t]
\centerline{\includegraphics[width=7cm]{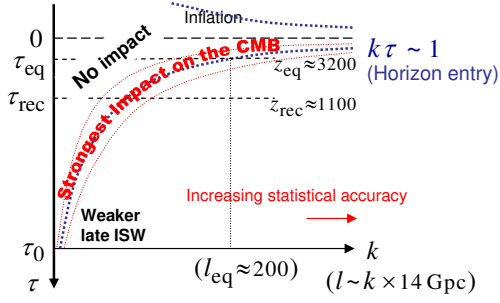}}
\caption{While the acoustic pattern of CMB temperature anisotropy
 is formed on the surface of last scattering ($z\rec\sim 1100$),
 the metric perturbations on this surface {\it do not play a major role\/}
 in the observed CMB spectra $C_\ell$.
 The potentials at last scattering do control the subdominant 
 ``baryon loading'' effect. Primarily, however,
 the CMB multipoles are sensitive to the gravitational potentials and underlying
 dark dynamics at the {\it horizon entry\/} of the corresponding modes 
 ($z\sim \ell^2/4$ for $z< z_{\rm eq}$,
  and $z\sim 20\,\ell$ for $z\gtrsim z_{\rm eq}$), 
 covering the entire range from $z\sim 1$ for lowest $\ell$
 up to $z\sim 6\times 10^4$ for $\ell\sim 3000$.}
\label{fig_hor_ent}
\end{figure}

\subsubsubsection{Overall}
The above results can be summarized concisely as follows.
At a given comoving scale $k$, probed by $\ell\sim kr \approx k\times 14$\,Gpc,
the CMB anisotropy is most sensitive to the value and evolution of
$\Phi+\Psi$ during the scale's horizon entry.
This is illustrated by Fig.~\ref{fig_hor_ent}.
Although most of the observed CMB photons scattered last during hydrogen recombination at
$z\rec\sim 1100$, 
the potentials during recombination are of relatively minor importance
for the observed anisotropies. 
(Exceptions are the scales of the first peak that happened to enter during recombination,
 and the nondegenerate baryon-loading signature on smaller scales.)

When the potentials do not decay quickly during the entry,
they significantly suppress the CMB temperature spectrum~$C_\ell$
(by a factor of 25 in the Einstein-de Sitter scenario with adiabatic perturbations.)
The $C_\ell$ response to changes of the potentials after the entry is weak.
After recombination, in particular, 
the $C_\ell$ response to the subhorizon changes of potentials
is diminished by a factor $\sim\Ha/k$.

In principle, $\Delta T/T$ of the CMB responds comparably well
to the metric inhomogeneities during the horizon entry at 
either high or low redshifts.
Nevertheless, the CMB  
is a considerably better probe of the horizon-scale potentials at high redshifts  
because of the reduced cosmic variance.
Let us define the wavevector $k_{\rm ent}$ of the CMB modes 
that enter at a redshift~$z$ by a condition
$k_{\rm ent}\,S(z)\equiv1$, where $S=\int c_s\,d\tau$ 
is the corresponding size of acoustic horizon.
The modes that enter at a redshift $z$ match to the multipole
$\ell_{\rm ent}=k_{\rm ent}\,r=r/S$.
Taking $S\approx \tau/\sqrt3$ as a valid estimate both
in the radiation era (when $R_b$ is negligible) 
and after recombination 
(when baryons decouple and formally $c^2_{\rm eff,\,\gamma}=1/3$),
we obtain 
\be
\ell_{\rm ent}(z)\equiv {r\over S}\approx 
 \lf\{\ba{l}
  z/18,~~~~~{\rm radiation\ era},\\
  \sim 2\sqrt{z},~~{\rm matter\ era},\\
 \ea\rt.
\lb{l_to_z}
\ee
estimated for a $\Lambda$CDM model with $\Omega_mh^2= 0.13$ and $h=0.7$.
For this model the redshift of equality $z_{\rm eq}\approx 3110$
and recombination $z_{\rm rec}\approx 1090$ match to 
$\ell_{\rm ent}\approx 200$ and $90$ respectively.

The 1-$\sigma$ uncertainty of probing a non-degenerate effect, quantified by a parameter $p$,
can be evaluated as \cite[e.g.,][]{KendallStuart2}
\be
\Delta p = 1\lf/\sqrt{\sum_l\lf(\fr{\pd C_l/\pd p}{{\rm r.m.s.}\,C_l}\rt)^2}.\rt.
\ee
For a CMB experiment limited on the studied scales only by cosmic variance,
for either temperature or polarization power spectra
$({\rm r.m.s.}\,C_l)/C_l=[(\ell+\fr12)f_{\rm sky}]^{-1/2}$\ct{Knox:1995dq,
                                                            Seljak:1996ti,Zaldarriaga:1997ch},
where $f_{\rm sky}\le1$ is the experimental sky coverage.
Then
\be
\Delta p = 1\lf/\sqrt{\sum_l(\ell+\fr12)f_{\rm sky}\lf({\pd \ln\, C_l\over \pd p}\rt)^2}.\rt.
\ee
Thus for constraining an effect that affects the modes which enter the horizon 
at a redshift~$z$ over a span of redshifts~$\Delta z$, by eq.\rf{l_to_z}, 
we can expect accuracy 
\be
\Delta p \sim  f_{\rm sky}^{-1/2}(\pd \ln\, C_l/\pd p)^{-1}\times 
               \lf\{\ba{l}
               18/\sqrt{z\,\Delta z},~~{\rm radiation\ era},~~~\\
               ~~1/\sqrt{2\,\Delta z},~~{\rm matter\ era}.
               \ea\rt.
\lb{eps_cmb}
\ee
This accuracy improves with increased $\Delta z$ and, when probing the radiation era, 
with increased $z$.

\section{Applications and caveats for modified gravity}
\label{sec_ModGrav}

We now briefly consider the possibility
that general relativity (GR) breaks down on cosmological scales.
Modified gravity (MG) offers rich phenomenology
by invoking new degrees of freedom whose dynamics is substantially different 
and often involves more parameters than the standard cosmological model.
It typically predicts nonstandard evolution of perturbations 
on horizon and subhorizon scales alike.

While the detection of such phenomena
as non-standard growth of cosmic structure or anomalous lensing
may indicate MG, we will see that 
without any restrictions on the dark dynamics,
the identical effects could always
be generated by a non-minimal dark sector
that influences the visible matter according to the standard 
Einstein equations.\footnote{
  We stress that these features remain useful hints of MG.
  Since they are even more ubiquitous 
  than the scenarios of modified gravity
  and may reveal other new physics,
  they are well worth searching for.
}
Further in this section we will argue that other features
should yet allow to discriminate MG observationally.

We will assume that even if full Einstein gravity
fails on cosmological scales,
the Einstein principle of equivalence remains valid for the visible species.
This assumption is common to many existing MG models.
It is motivated by the relatively strong terrestrial and solar-system 
constraints on the equivalence principle.

Thus we suppose that the regular matter couples covariantly 
to a certain matter-frame ``physical'' metric~$g_{\mu\nu}$.
However, we now neither take for granted that all dark fields
also couple covariantly to the same metric~$g_{\mu\nu}$,
nor assume that the dynamics of $g_{\mu\nu}$ itself
is governed by the Einstein equations.

Under the weaker assumption of the equivalence principle 
for only the visible matter, 
all observable signatures
of new physics can still be quantified
by {\it any of the three parameterization schemes\/} of Sec.~\ref{sec_dynamics}.
Indeed, the  $g_{\mu\nu}$ background can still be described
by a single number for its present spatial curvature  
and by its uniform redshift-dependent expansion rate $\Ha(z)$.
The potentials $\Phi$ and $\Psi$,
defined by eq.\rf{Newt_gauge} to
parameterize the inhomogeneities of the physical metric $g_{\mu\nu}$,
will play their usual role in the evolution of light and baryons.
Moreover, the (effective dark) energy and momentum densities
assigned to the missing sources of curvature
by the naive application of the Einstein equations
will evolve in agreement with the usual local conservation laws,
which is easily seen as follows.

Let by definition 
\be
T^{\mu\nu}_{\rm eff~dark}\equiv 
       \fr1{8\pi G}\,G^{\mu\nu}-\sum_{{\rm known}~a}T^{\mu\nu}_a,
\lb{T_eff_dark}
\ee
where the last sum is over the known regular particles.
Their energy-momentum tensor 
[constructed unambiguously from the species' action $S_a$ as
 $T^{\mu\nu}_a=(2/\sqrt{-g})\,\,\delta S_a/\delta g_{\mu\nu}$]
is covariantly conserved by the assumed covariance for the regular species.
The matter-frame Einstein tensor $G_{\mu\nu}=R_{\mu\nu}-\fr12g_{\mu\nu}R$,
where $R_{\mu\nu}$ is the Ricci tensor of the physical metric,
is also covariantly conserved by the Bianchi identities.
Thus the entire expression\rf{T_eff_dark} is covariantly conserved:
\be
T^{\mu\nu}_{\rm eff~dark\,;\nu}=0,
\lb{eff_conserv}
\ee 
with all covariant derivatives being
taken relative to the physical metric.

Since $T^{\mu\nu}_{\rm eff~dark}$ is covariantly conserved,
the background and perturbations 
of the missing energy and momentum evolve according to  
equations\rf{dot_rho}\,--\rf{dot_v}, derived from the identical conservation law.
Thus if all our probes of the invisible degrees of freedom 
are based solely on their gravitational impact on light, baryons, 
and other regular particles 
(neutrinos, WIMP's when probing dark energy, etc.)
then phenomenologically 
{\it all observable signatures\/} of a considered MG model 
can be mimicked by an effective GR-coupled dark sector.
Specifically, we can find the corresponding effective 
$w(z)$ to reproduce the missing energy background,
and the effective anisotropic stress $\sigma(z,k)$ and stiffness
$c^2_{\rm eff}(z,k)$ to describe scalar perturbations.
For example, for a popular DGP gravity model\ct{DGP00}
this was demonstrated explicitly by\ctt{KunzSapone06}.

With this discouraging general conclusion, 
we may enquire whether cosmology
can at all reveal definitive distinctions between MG and 
general relativity with a peculiar yet physically permissible dark sector.
To establish such distinctions,
let us summarize the conceptual differences of MG
from general relativity:
\begin{itemize}
\item[A.]
Some dark degrees of freedom may not couple covariantly to the matter metric $g_{\mu\nu}$.
\item[B.]
The gravitational action may not be given entirely by 
the Hilbert-Einstein term
$S_{\rm grav}=(16\pi G)^{-1}\!\int d^4x\sqrt{-g}\,R$.
\end{itemize}
The distinctive observable consequences of these special properties of MG
may include:
\begin{itemize}
\item[1.]
The effective dark dynamics, 
which is observationally inferred by assuming the Einstein equations,
violates the equivalence principle (EP).
The EP violation can be seen, e.g., as 
 \begin{itemize}
\item[i.]
The dependence of the inferred local dark dynamics
on the distribution of visible matter,
when it cannot be explained by non-gravitational 
dark--visible coupling allowed by particle experiments. 
\item[ii.]
Superluminality of the inferred dark dynamics.
 \end{itemize}
\item[2.]
The dynamics of gravitational waves (tensor modes) deviates from the predictions
of the Einstein equations,
assuming that both the visible and inferred dark species
contribute to the energy-momentum tensor in the simplest way.
\end{itemize}
Most of these signatures have already been utilized for falsifying
MG models with existing or suggested 
observations,\ct{Clowe06ApJL,Bradac:2006er} 
and\ct{KahyaWoodard07}; we comment additionally on them next.

\subsection{EP violation for the inferred dark dynamics}
\label{sec_MG_scal}

The violation of the first condition
can be illustrated by an extreme toy theory in which 
the regular matter (``baryons,'' for short)
constitute all independent degrees of freedom.
Let the metric in this theory be specified 
by baryon distribution via some deterministic relation
(e.g., as $g^{\mu\nu}=\Lambda^{-1} T^{\mu\nu}_{\rm baryon}$,
 following from an action 
$S=S_{\rm baryon}-\int\!d^4x\,\sqrt{-g}\,\Lambda$
with $\Lambda$ being a constant).
Even in such a contrived theory,  by the above arguments,
the effective missing energy and momentum densities\rf{T_eff_dark}
would appear to evolve and gravitate 
in agreement with energy-momentum conservation and the Einstein equations.
In this example, however, the effective dark density and stress are
uniquely determined by the distribution of the visible matter.
This does not occur for truly independent dark degrees of freedom.

In more realistic MG theories 
we should not expect a deterministic relation between
the visible and effective dark distributions.
Still, if the dark and visible sectors interact
other than by coupling covariantly to the common metric
then the inferred laws of the effective local dark dynamics
would depend on the visible environment.

Detection of such dependencies would be particularly feasible for the dark matter,
for which there are plentiful observable regions with varying environment:
varying in both visible matter density
and in its ratio to dark density.
In addition to baryons clumping strongly at low redshifts,
the density of all known visible species 
varies with redshift due to the Hubble expansion.
The ratio of visible to dark density is perturbed even on large scales
due to dark matter 
being decoupled from the photon-baryon acoustic oscillations 
and clustering on its own until recombination.
The segregation of dark matter and baryons is even more apparent
at late times on the scales of clusters and smaller,
becoming almost complete in the famous bullet cluster 
example\ct{Markevitch:2003at,Clowe06ApJL,Bradac:2006er}.

In order to test that the standard CDM+GR model accounts for the signatures
of dark matter at various redshifts and scales,
it is crucial to observe and robustly model 
the dynamics on a wide range of scales, including the highly non-linear ones.
It is also important to utilize complimentary probes, such as
baryonic matter (responding to $\Phi$),
and the CMB or gravitational lensing (probing $\Phi+\Psi$). 
Various signatures of dark matter 
that allow to compare dark matter parameters under different conditions
include: 
the height alteration of the odd and even CMB acoustic peaks
due to the CDM potential affecting the coupled baryons
before recombination\ct{HuSugSmall96},
the early ISW enhancement of the first peak
\citep[e.g.,][]{HuNature},
the significant suppression of the CMB temperature power $C_l$
below the first peak 
(Sec.~\ref{sec_cl_resp_quant} and Fig.~\ref{fig_CMB_mat}),
the linear and nonlinear dynamics of cosmic structure,
and the lensing of the CMB and background galaxies by the 
CDM potential at lower redshifts.

Dark energy (DE) 
does not trace any visible species at low redshifts,
when its density appears almost redshift-independent.
DE may track the radiation or matter background at higher redshifts; 
then its perturbations are not expected to evolve similarly
to any of the standard species. 
(For example, see the evolution of the perturbations of tracking quintessence
in the radiation and matter eras on the right panels of 
Figs.~\ref{fig_CMB_rad} and~\ref{fig_CDM} respectively.)
Constraints on the background equation of state $w$
can rule out specific DE or MG models 
but do not decisively differentiate between the DE and MG paradigms.
Beyond~$w$, the perturbations of dark energy are likely to 
be stiff and thus can be constrained only during horizon entry.
As discussed previously, the CMB can provide such constraints
over a wide range of redshifts, from $z\sim 1$ up to $z\sim 10^5$,
becoming much tighter toward higher redshifts
(Sec.~\ref{sec_impacts} and Fig.~\ref{fig_hor_ent}).

With the cosmological constant fitting the current data well,
the search for {\it any manifestation\/} of non-trivial DE or MG dynamics 
may be the utmost priority for establishing the origin of cosmic acceleration.
This includes falsifying the background equation of state $w(z)\equiv -1$ at low redshifts
but by no means is limited to $w(z)$.
Examples of other searches, which may end up being more fruitful, are:
the search for a subdominant non-standard component
in the radiation or early matter era 
(Secs.~\ref{sec_stiffness} and \ref{sec_clustering}),
nonstandard growth of cosmic structure\ct{Linder_gamma05},
and nonstandard $\Phi/\Psi$\ct{Bert06},
probed by comparing the CMB or lensing signal
with the galaxy and cluster distributions.

\subsection{Superluminal dark flows}
\lb{sec_MG_superlum}

Another possible signature of MG 
is the superluminal propagation of dark
perturbations.
The superluminality may be spurious,
from our describing MG
by a parameterization that assumes general relativity.
It may also be real, 
from the equivalence principle not applying to some dark degrees of freedom.

While numerous observational tests for superluminality can be thought of, 
here we consider only one specific example.
Spurious or real superluminal propagation
of (effective) dark inhomogeneities can be constrained by the sensitivity of
the CMB acoustic phase to the propagation velocity of dark perturbations 
(Sec.~\ref{sec_speed}).
A straightforward Fisher-matrix forecast shows that
TT, TE, and especially EE spectra from the future high-$\ell$ CMB experiments, 
Planck\footnote{http://www.rssd.esa.int/Planck} and 
ACT\footnote{http://www.physics.princeton.edu/act} in particular,
will strongly restrict the abundance of any dark component
that supports scalar perturbations with $c_p^2>1/3$, including $c_p>1$.

If $B$-polarization of the CMB reveals the signal
of relic gravitational waves\ct{SelZald_Bmode,Kamionk96_Bmobe}, 
a similar effect in the tensor sector  
will directly constrain the streaming dark species for which 
streaming velocity $c_p$ exceeds the speed of gravitational waves, $c_{s,\,\rm grav}$.
[Indeed, $c_{p,\,\nu}>c_{s,\,\rm grav}$, 
albeit under somewhat different conditions,
for the dark matter emulator of \ctt{KahyaWoodard07}, 
 discussed below.]
The oscillation amplitude of the gravitational (tensor) modes
after horizon entry
is noticeably affected by neutrino perturbations\ct{Weinberg:2003ur}.
However, the phase shift of the tensor oscillations
is strictly forbidden\ct{SB_tensors05} when for all species 
$c_{p,\,\rm dark}\le c_{s,\,\rm grav}$, as required by GR.
If, on the other hand, the velocity of neutrinos
or other dark species with non-negligible abundance  
exceeds $c_{s,\,\rm grav}$
then the arguments of\ctt{SB_tensors05} show that 
the phase of the tensor BB signal in the CMB 
will be necessarily shifted.

\subsection{Nonstandard phenomenology of gravitational waves}
\label{sec_MG_tens}

An interesting test of a broad class of MG
alternatives to dark matter 
was recently suggested by\ctt{KahyaWoodard07}. 
They considered the propagation of a fundamental tensor field $\tilde g_{\mu\nu}$
[whose kinetics is governed by 
$S_{\rm grav}=(16\pi \tilde G)^{-1}\!\int d^4x\sqrt{-\tilde g}\,\tilde R$]
in any of the model where $\tilde g_{\mu\nu}$ 
is sourced by the luminous matter alone 
by the Einstein law
\be
\tilde G^{\mu\nu}\approx 8\pi G T^{\mu\nu}_{\rm luminous}.
\lb{tildeEeq}
\ee
The physical metric $g_{\mu\nu}$ was set to reproduce the observed
gravitational potentials in the vicinity of our galaxy.
The knowledge of the other specifics of the MG dynamics was therefore unnecessary. 
Then\ctt{KahyaWoodard07} showed that the arrival of the gravitational waves 
from a cosmological event, e.g.\ a supernova, is noticeably {\it delayed\/}
with respect to the arrival of the associated neutrinos and photons.

While the assumption\rf{tildeEeq}, implicit in\ctt{KahyaWoodard07}, 
incorporates many models it is not generic.  
For example, it does not apply if 
\be
\tilde G^{\mu\nu}\approx 8\pi \tilde G T^{\mu\nu}_{\rm luminous},
\ee
where $\tilde G$ differs from the local value of
Newton's gravitational constant $G$.
Even in the TeVeS model\ct{Bekenstein04}, 
motivating\rf{tildeEeq},
$\tilde G\not=G$ strictly,
although it is natural to assume that for TeVeS
this difference is small\ct{Bekenstein04}.
Yet more generally, $\tilde G$ may 
depend on the new MG degrees of freedom, e.g.,
on the scalar field $\phi$ of tensor-scalar or 
tensor-vector-scalar models.

In any case, even if eq.\rf{tildeEeq} fails and the
exact prediction for the gravity wave delay by\ctt{KahyaWoodard07}
does not apply,
the gravitational waves and neutrinos or photons would still generally propagate 
with different velocities.
Thus the future gravitational wave astronomy can offer
robust tests for discriminating GR-coupled dark matter from modified gravity.

\section{Summary}
\label{concl}

\subsection{Approach}

Constraints on the {\it inhomogeneous\/} dynamics of the dark sectors
are essential to full understanding of their nature.
Varying with both redshift and {\it spatial scale\/}~$k$, 
the observable imprints of dark perturbations provide plentiful information
about the dark sectors' local kinetic properties and (self) interactions.
This information significantly complements the dark species' background
equation of state $w(z)$.

Reliable extraction of the information encoded in the dark perturbations
is hindered by the indirectness of their, sometimes subtle, 
gravitational impact on observables.
It is also obstructed by the numerous contributions to the observables 
from other complex multiscale cosmological and astrophysical phenomena.    
Moreover, controlled repeatable measurements of dark sectors 
cannot be afforded when the measuring device is the entire 
observable large-scale universe.
Nevertheless, the unavoidable ``nuisance'' phenomena and contaminations 
can be counteracted by detailed understanding of the involved physics 
and by complementary probes of as many independent characteristics
of the dark sectors as can be observationally accessed.

For a tractable and systematic study of the dark sectors
beyond the background equation of state,
we map the local kinetic properties of inhomogeneous dark dynamics at a given redshift to 
the characteristics of the observed cosmological distributions
that can reflect those dark properties. 
The correct mapping is more likely to be achieved with
a description of evolution that is simple 
and that manifests the objective causal relations explicitly.

The dynamics of gravitationally coupled perturbations of dark and visible species 
during horizon entry is pivotal to probing the dark inhomogeneities.
Then any dynamical species, including dark radiation and 
all dark energy candidates with $w\not\equiv-1$, are necessarily perturbed
and leave imprints of their highly specific inhomogeneous kinetics on the observable probes.
The contribution of subhorizon perturbations of a given magnitude $\delta\rho/\rho$ 
to gravitational potentials is suppressed, as $\Phi \sim (\Ha/k)^2\,\delta\rho/\rho$.

On the scales of the horizon and beyond, the apparent gravitational impact 
of the dark species on the visible ones depends strongly on the description used.
In most descriptions, the apparent inhomogeneities of the visible species
are changed by the properties that dark species have
long {\it before\/} and even long {\it after\/} the change. 
This misguides the identification of the observable features 
that reflect the internal dark properties at specific epochs.

The apparent cause--effect mismatch 
is, nevertheless, not intrinsic to linearly perturbed cosmological evolution.
Within Einstein gravity,
the internal local properties of dark species at a past time~$\tau$ 
affect only the perturbations
that had approached or entered the horizon by the time~$\tau$.
The mapping of dark properties to evolutionary changes
is unambiguous after horizon entry, 
when baryonic and photon perturbations
are instantaneously impacted by the Newtonian potentials,
then reflecting the instantaneous overdensity of visible and dark species.
The remaining, confined to the horizon entry, gravitational impact
of the dark sectors at a given~$\bk$ in linear theory
is also unambiguous (Sec.~\ref{sec_uniq}).

%\footnote{
% A perturbation whose change is considered since the horizon entry
% can be quantified naturally while it is beyond the horizon: 
%Before the perturbation has entered and remains frozen, 
%we can impose dynamically the FRW metric, 
%in which the quantification is unambiguous.
%Within linear theory, the result does not depend on how the FRW metric is imposed 
%and this is, in principle, always possible \cite[for details, see][]{B06}.
%}

It is easy to describe the full perturbed
linear cosmological dynamics
(including that of partly polarized photons, baryons,
realistic neutrinos, quintessence, and other particles or fields)
by a formalism
in which the changes of perturbations in the visible sectors 
are concurrent with the local dark properties responsible for these changes\ct{B06}.
The resulting description reveals explicitly the objective causal dependencies
and enables us to map observable features to local dark properties
more reliably than with traditional formalisms, which lack this concurrence. 
The suggested formalism considers perturbations of canonical rather than proper distributions.
It has additional useful technical benefits,
for example, note (A) and (B) in the caption of Fig.~\ref{fig_CMB_analogy},
illustrating the corresponding description of the acoustic CMB modes.

Using this  formalism, we relate general properties of dark perturbations
to observable features by tracking the gravitationally coupled evolution 
of dark and visible perturbations.
The advantages of such an evolutionary study 
over black-box computation of observables are twofold:
It isolates all observable signatures of the studied phenomena;
it also reveals the mechanisms that generate these signatures
and allows us to judge the mechanisms' robustness.

\subsection{Sensitivity of probes}

We categorize the primary probes, 
responding to the dark dynamics through linear evolution,
as either ``light'' or ``matter'', Sec.~\ref{sec_probes}.
Those of the first type (CMB spectra) probe the trajectories close to null geodesics;
of second type (matter transfer functions) respond to the metric along 
the time-like Hubble-flow worldlines.

Primary CMB anisotropies are highly sensitive to the values and evolution of 
the potentials {\it during horizon entry\/}.
They are only mildly affected by 
the gravitational potentials on subhorizon scales.
(The situation may differ for secondary, nonlinear, CMB features.)
For example,
after recombination the response of the CMB angular power spectrum $C_\ell$  
to a change of subhorizon potentials
is easily quantifiable with eq.\rf{Cl_resp_late}, derived in the Limber approximation.
This response is suppressed relative to the contribution\rf{Cl_late_ent_inv}
from the horizon entry by a factor $\fr{r}{\ell\delta\tau}$.
By the arguments of Sec.~\ref{sec_cl_resp_quant},
for linear changes of subhorizon potentials 
this factor is much smaller than unity ($\sim \Ha/k$).

The CMB is an excellent probe of 
the potentials on the horizon scales at high redshifts.
The dark inhomogeneities that enter the horizon at a redshift~$z$
affect most of all the CMB multipoles with $\ell\sim z/20$ for the radiation epoch
and $\ell\sim 2\sqrt{z}$ for the matter era, eq.\rf{l_to_z}.
With sufficient angular resolution of the detector 
and reliable subtraction of foregrounds and secondaries, 
the larger number of statistically-independent
multipoles at higher $\ell$'s
improves the constraints on the parameters 
that describe the dynamics at higher redshifts
$(z,\,z+\Delta z)$ as $\sqrt{z\,\Delta z}$ for radiation
and $\sqrt{\Delta z}$ for matter domination, eq.\rf{eps_cmb}.

In contrast, matter transfer functions 
are more sensitive to the potential $\Phi$ at low redshifts.
The gravitational impact on massive matter at the later times is 
erased less by the Hubble friction.
The response of matter overdensity to $\Phi(\tau)$
at any past epoch since the horizon entry
can be quantified by a simple equation
$\Delta (\delta\rho_m/\rho_m)_{\rm today}\sim  - \,\tau\Delta\tau\,k^2\,\Phi(\tau)$,
eq.\rf{growth_grf}.
The reduction of the matter sensitivity to early-time dynamics
is manifested in the prefactor~$\tau\Delta\tau$.

\subsection{Mapping the dark properties}

The potentially measurable properties of arbitrary dark sectors
may be parameterized by a single function of $z$ for background
and two (transfer) functions of $z$ and $k$ for scalar perturbations.  
The effects of dark sectors or modified gravity on the metric 
may be phenomenologically described in terms of
$\{\Ha$, $\Phi$, $\Psi\}$.
Assuming the Einstein equations,
we may instead consider either dark species' densities and stresses
$\{\rho$, $\delta^{(c)}$, $\sigma\}$
or their dynamical characteristics $\{w$, $c^2_{\rm eff}$, $\sigma\}$ 
(all reviewed in Sec.~\ref{sec_dynamics}).
The correspondence between the inhomogeneous properties of the dark universe
and the observable characteristics of the CMB and LSS
is summarized by Table~\ref{tab_sum}
and the text that follows.

\subsubsection{$\sigma$ vs. $c_{\rm eff}$}

The measurable dynamical characteristics of 
scalar dark {\it inhomogeneities\/} may be fully quantified 
by the anisotropic stress potential 
$\sigma$, eq.\rf{sigma_def},
and stiffness (``effective sound speed'') 
$c^2_{\rm eff}$, Sec.~\ref{sec_param_dyn}.

For adiabatic initial conditions, 
the gravitational potentials $\Phi(\tau,k)$ and $\Psi(\tau,k)$
depend on $c^2_{\rm eff}$ only at the order $(k\tau)^2$,
i.e., at a relatively late stage of horizon entry, Sec.~\ref{sec_stiffness}.
On the contrary, anisotropic stress of streaming species
affects the potentials earlier, already at the order $(k\tau)^0$,
Sec.~\ref{sec_anis_stress}\ct{MaBert95}.
This distinction may be given the following intuitive illustration:
The stiffness affects the motion of the dark species,
hence a certain time is required for the dark species to redistribute  
and start sourcing different potentials.
On the other hand, anisotropic stress is generated 
without displacing the matter.
Being itself a source of curvature in the Einstein equations,
anisotropic stress changes the gravitational potentials earlier.

There are several observable consequences 
of the early and late influence respectively of $\sigma$ and $c^2_{\rm eff}$:
Both $\sigma$ of freely streaming relativistic species and
$c^2_{\rm eff}$ of a component of dark radiation as stiff as quintessence
($c^2_{{\rm eff},\,\phi}=1$) reduce $\Phi+\Psi$.
Yet the corresponding reductions occur at a different phase of the acoustic CMB oscillations.
As a result, the CMB oscillations in the radiation era are somewhat suppressed
by the gravitational effect of the anisotropic stress of streaming neutrinos,
yet are slightly boosted by the gravitationally coupled
perturbations of tracking quintessence, Fig.~\ref{fig_CMB_rad}.\footnote{
 The example of tracking quintessence shows that the reduction 
 of the magnitude of dark perturbations
 {\it does not necessarily imply\/} 
 damping of the gravitationally coupled CMB fluctuations, 
 as sometimes stated.  
 The sign of the impact on the CMB anisotropy depends 
 not only on the sign of the change of $\Phi+\Psi$, 
 sourced by the dark perturbations, 
 but also the {\it time\/} of this change.
}

Another consequence of the lateness of the gravitational impact of~$c^2_{\rm eff}$  
is a considerably larger ratio of the matter response over the CMB response 
to a change of~$c^2_{\rm eff,\,dark}$,
as compared to this ratio for $\sigma_{\rm dark}$.
These results, summarized on the upper half of Table~\ref{tab_sum}, 
should help distinguish an excess of relic relativistic particles from
a subdominant tracking classical scalar field in the radiation era.

\begin{table*}[t]
\begin{center}
\caption{\lb{tab_sum}}
\vspace{-.1cm}
{\renewcommand{\arraystretch}{0}
\begin{tabular}{|c|clll|}
\hline
\rule{0pt}{2pt}& & & & \\
\hline
\strut      &  Quantified  &  ~Important  &  ~~Effect on   & ~~~~Effect on  \\
\strut\raisebox{1.4ex}[0pt]{Property} 
            &       by     &   ~~~~~~for  &  ~~the CMB    &  ~~~~~~Matter   \\
\hline
\rule{0pt}{2pt}& & & & \\
\hline
\strut      &              &    Early    &  Amplitude   & Minor on power \\
\strut Anisotropic & ~~$\sigma$,~ eq.\rf{sigma_def} 
\strut      &  stage of   & (Suppressed   & (Enhanced       \\
\strut  Stress  &[$\Phi-\Psi$,~ eq.\rf{Psi-Phi}]~~
            &   horizon   & ~by $\sigma$ from & ~by $\sigma$ from \\
\strut      &             &   entry   &  ~streaming)   & ~streaming)     \\
\hline
\strut      &            &   Late      &  Amplitude    &  Medium on power  \\
\strut      &            &  stage of   & (Enhanced~~  & (Suppressed     \\
\strut\raisebox{1.4ex}[0pt]{Stiffness}
            & \strut\raisebox{1.4ex}[0pt]{$c^2_{\rm eff}$,~ eq.\rf{c_eff_def}}
                               &  horizon  &  ~by tracking    & ~by tracking  \\
\strut      &                  &  entry    &  ~quintessence) &  ~quintessence) \\
\hline
\strut Velocity of &           & Features  & Phase of   & Phase of \\
\strut a perturbation 
            & $c_p$,~ Sec.~\ref{sec_speed}
                               &  local in  &   the acoustic     & baryonic  \\
\strut front &              &  real space    &   peaks      &  oscillations  \\
\hline
\strut      &             & Horizon              &  Significant  &    Primary       \\
\strut    & $\Phi$, $\Phi+\Psi$ & entry (CMB)    &  suppression  &    driving of    \\
\strut\raisebox{1.4ex}[0pt]{Self-clustering}
          & eq.\rf{Newt_gauge}&  and subhorizon  &    of the     &    the structure \\
\strut            &       & evolution (LSS)      &   amplitude   &     growth       \\
\hline
\vspace{-1.2cm}
\tablecomments{
         Summary  of the discussed properties of the dark sectors,
         the epochs of their observational relevance,
         and their effects on the CMB power spectra and 
         on large-scale structure.}
\end{tabular}
}
\end{center}
\end{table*}

\subsubsection{$c_{\rm eff}$ vs. $c_p$}
\lb{sec_sum_cp}

Two other potentially observable characteristics of the dark species 
that should be distinguished from each other are the species'  
$c_{\rm eff}(k)$, considered above,
and the velocity $c_p$ of the wavefront of their localized perturbation.
These quantities need not be functionally related.\footnote{
 Since $c^2_{\rm eff}(z,k)$ and  $\sigma(z,k)$ provide the most general parameterization 
 of the observable dynamical properties of the dark perturbations, 
 these functions jointly should encode all the observable signatures of $c_p$.
 Nevertheless, since $c_p$ has an important kinetic meaning and maps to 
 a sharp non-degenerate signature in the CMB spectra,
 it is worthwhile to consider this quantity on its own,
 and to compare it with $c_{\rm eff}(k)$.
}
Together, they are powerful indicators of the nature of dark sectors.
For example, interacting relativistic particles whose 
free-flight time is much smaller than the Hubble time have
$c_{\rm eff}(k)=c_p=1/\sqrt3$;
free-streaming relativistic particles have
$c_{\rm eff}(k)=1/\sqrt3$ while $c_p=1$;
quintessence is characterized by
$c_{\rm eff}(k)=c_p=1$.

Observationally, the phase of the CMB acoustic oscillations
in both temperature and $E$-polarization power spectra or their 
cross-correlation is shifted
if and only if $c_p$ of any of the dark component
in the radiation era exceeds the acoustic sound speed $c_s\simeq 1/\sqrt3$,
Sec.~\ref{sec_speed}\ct{BS04}.
Any such dark component contributes to the phase shift by an easily calculable amount. 
Importantly, the additive shift of the acoustic peaks  for adiabatic perturbations
is nondegenerate with any of the standard cosmological parameters
or with the shape of the primordial power spectrum\ct{BS04}.

Thus for the robust knowledge of the nature and kinetics 
of the dark radiation
(encompassing neutrinos, possibly other light particles, and early quintessence)
both $c_{\rm eff}$ and $c_p$ 
should be targeted by experimental strategies and data analyses.
In particular, the determination of $c_{\rm eff}$ appears most promising from
the comparison of the {\it CMB and LSS\/} power.
On the other hand, $c_p$ is probed increasingly best by the CMB spectra 
extended toward {\it higher\/} $\ell$'s, with the {\it polarization\/} autocorrelation (EE)
being the most crucial\ct{BS04}.

\subsubsection{$\Phi$ and $\Phi+\Psi$}

We can probe the inhomogeneous properties of the invisible universe
more model-independently by constraining
metric perturbations directly. 
Such constraints have a greater validity than general relativity, 
implied for deducing the dark dynamical properties. 
They can be placed meaningfully
under a weaker assumption of only the visible sectors  
obeying the equivalence principle,
better constrained for them by terrestrial and solar-system probes.

In this perspective, the discussed probes of the 
scalar component of anisotropic stress~$\sigma$ can be viewed as 
the direct probes of the difference of the Newtonian-gauge potentials 
$\Phi-\Psi$, eq.\rf{Psi-Phi}.
The primary anisotropies of the CMB 
are a sufficiently clean probe of the sum $\Phi+\Psi$,
entirely responsible for the gravitational driving  
of the scalar CMB inhomogeneities on all scales and epochs
except for a narrow band of redshifts 
around $z_{\rm dec}\sim 1100$.
CDM, on the other hand, responds strictly
to the Newtonian potential~$\Phi$ on all scales and at all times,
and so does the baryonic matter on large scales after 
baryon decoupling from the CMB.

Trivially [eq.\rf{d_c_sol}],
the matter growth function is almost directly proportional
to past $\Phi(\tau)$ (as previously noted, weighted by its conformal time~$\tau$.)
The CMB is also affected by the gravitational potentials.
Delay of the decay of $\Phi+\Psi$ after the horizon entry
suppresses the CMB power by over an order of magnitude.
[In CDM-dominated limit by $5^2=25$ times\ct{B06} ---
more than by a factor $2^2=4$ which appears
in the apparent description of the Sachs-Wolfe effect 
in terms of the proper Newtonian perturbations.]

This suppression of the low CMB multipoles is physical:
It is absent in models where the metric in the matter era is unperturbed.
The suppression is diminished in the models in which
the metric is perturbed less than in the $\Lambda$CDM scenario.
The prominent suppression of the CMB power at $l\lesssim100$ by the CDM 
potential is one of the primary reasons that  
models without dark matter provide poor fits to the CMB data.
The suppression of the CMB temperature anisotropies for $\ell<100$ 
must not be explained as a ``resonant self-gravitational driving'' 
of radiation perturbations at $l\gtrsim200$:
This effect is caused by and probes 
the inhomogeneities of matter after equality and 
not the inhomogeneities of radiation before equality.

The suppression of $C_\ell$ for $\ell\lesssim 100$
severely restricts the alternatives to CDM and the models of dark energy 
which reduce metric perturbations at any redshift in the matter era.
Examples of such mechanisms are contribution of quintessence 
to the density in the matter era,
interaction or unification of dark matter and dark energy
\citep[e.g.,][]{Wetterich88,PerrottaBaccig02,
FarrarPeebles04,Catena:2004ba,Scherrer_kess}, 
or MOND-inspired alternatives to dark-matter\ct{Milgrom83,Bekenstein04}.

\subsection{Modified gravity}
\lb{sec_sum_MG}

Many authors have suggested that 
modification of general relativity on cosmological scales
is the cause of the cosmic acceleration 
\citep[for recent reviews see][]{Copeland:2006wr,Nojiri:2006ri}
or even of the apparent manifestations of dark matter\ct{Milgrom83,Bekenstein04}. 
In Sec.~\ref{sec_ModGrav} we consider the phenomenology of 
typical models of modified gravity (MG) 
that retain the equivalence principle for the visible sectors. 
We show that in these models all gravitational impact of the hidden physics 
can be described within the same parameterization schemes 
of Sec.~\ref{sec_dynamics},
developed to quantify the observable properties of dark sectors
that are coupled by general relativity (GR).
Indeed, these schemes were restricted only by the covariance of the visible dynamics,
the assumption of the Einstein equations,
and the local conservation of the dark energy and momentum. 
However, for any covariant visible dynamics, 
the formal dark energy-momentum tensor\rf{T_eff_dark} 
that is missing in the Einstein equations
is covariantly conserved automatically (Sec.~\ref{sec_ModGrav}).
Thus all observable signatures of MG can be mimicked by
effective dark energy and momentum that
influence the visible species according to the Einstein equations
and during evolution are conserved locally. 

Particularly, the nonstandard structure growth or $\Phi/\Psi$
ratio that are predicted by {\it any\/} MG model of the considered
broad class can in principle be reproduced without violation of
the Einstein equations by sufficiently peculiar dark dynamics.
Nevertheless, first, such signatures would still signal some
non-minimal physics and therefore should be considered for
experimental constraints whenever possible.  Second, GR remains
falsifiable by other effects; in particular, by
the violation of the equivalence principle by apparent dark dynamics.  
This may be manifested in the strong dependence of the dark dynamics on the
visible matter (Sec.~\ref{sec_MG_scal}), and in the signatures of
effective or real superluminal dark flows (e.g.,
Sec.~\ref{sec_MG_superlum}).  GR can also be falsified by nonstandard
phenomenology of gravitational waves \citep[e.g.][and
Secs.~\ref{sec_MG_superlum} and~\ref{sec_MG_tens}]{KahyaWoodard07}.

~

~

\section*{Acknowledgments}
I am grateful to Salman Habib and Katrin Heitmann for
valuable discussions, suggestions, and comments on the manuscript.
I thank  Daniel Holz and  Gerry Jungman for stimulating talks and 
useful suggestions.
This work was supported by the
US Department of Energy via the LDRD program of Los Alamos. 

\appendix

\section{CMB sensitivity to $\delta\Phi$ on small scales}
\lb{sec_appx}

We quantify the CMB sensitivity
to the metric inhomogeneities on subhorizon scales and at late times
using the Limber approximation\ct{Limber54},
applied to the CMB by 
\citet{Kaiser84,Kaiser92} and \citet{HuWhite_weak_coupling95}.
The Limber approximation for the CMB power spectrum $C_l$
assumes that the change of the source~$S$ in the line-of-sight integral\rf{los_int}
is negligible over the wavelength of a typical contributing mode $k^{-1}\sim r/\ell$.
If $\delta \tau$ is a temporal scale over which the source 
changes by an order of unity then the above condition is equivalent to 
$k\,\delta \tau\gg 1$.
In the Limber limit, $C_l$ is primarily contributed by the modes with  
$|\bk\cdot\bn|\,\delta \tau \lesssim 1$, while
positive and negative contributions to $\Delta T/T$
from the peaks and troughs of the other modes cancel\ct{HuWhite_weak_coupling95}.

The ISW contribution to the anisotropy source\rf{los_source} is
\be
S_{\rm ISW}=g\,(\dot\Phi+\dot\Psi).
\ee
This direction-independent source in the Limber approximation
gives the following additive contribution 
to the power spectrum\ct{Kaiser92,Hu:1999vq}: 
\be
\delta C_\ell\approx \fr{2\pi^2}{\ell^3}\int dr\ r\ 
                       \Delta^2_{\rm ISW}(k=\ell/r),
\lb{Cl_ISW}
\ee
where $\Delta^2_{\rm ISW}(k)\equiv k^{3}P_{\rm ISW}(k)/(2\pi^2)$
is the dimensionless power spectrum of $S_{\rm ISW}(\bk)$,
evaluated at a time $\tau(r)$. 
Thus the ISW contribution is
\be
\delta C_\ell\approx \fr{2\pi^2}{\ell^3}\int dr\ r\ 
    g^2(\dot\Phi+\dot\Psi)^2\ \Delta^2_\zeta(k=\ell/r),
\lb{Cl_ISWt}
\ee
where $\Phi$ and $\Psi$ now denote the potentials' transfer functions,
i.e., the potentials in the perturbation modes that are normalized to a unit
primordial curvature perturbation, $\zeta_{\rm in}(\bk)\equiv 1$.

We now quantify the sensitivity of 
the power spectrum to a change of the potential 
$\delta \Phi$ on subhorizon scales over a time interval 
$\delta\tau\gg r/\ell$. For an estimate, in eq.\rf{Cl_ISWt} we set 
$g\approx 1$ at the considered late times, 
neglect the variation of $\Delta^2_\zeta(\ell/r)$
for the nearly scale-invariant primordial power,
and evaluate the remaining integral as
\be
\int dr\ r\ (\dot\Phi+\dot\Psi)^2 \sim r\,\fr{[\delta(\Phi+\Psi)]^2}{\delta\tau}.
\nonumber
\ee
Then    
\be
\delta C_\ell \sim \fr{2\pi^2\Delta^2_\zeta(\ell/r)}{\ell^2}\
                   [\delta(\Phi+\Psi)]^2\, \fr{r}{\ell\delta\tau}  .
\lb{Cl_resp_late_appx}
\ee

\bibliographystyle{apj}                       %% AASTeX
\bibliography{dk_bib}
%\vfill

\end{document}